\PassOptionsToPackage{table}{xcolor} 

\documentclass[manuscript,screen]{acmart}
\usepackage{geometry}
\usepackage{fancyhdr}
\usepackage{todonotes}
\usepackage{amsmath}
\usepackage{mathtools}
\usepackage{listings}
\usepackage{tcolorbox}
\usepackage{hyperref}
\usepackage{tabularx}
\usepackage[capitalise]{cleveref}
\crefformat{section}{#2\S#1#3} 
\crefformat{subsection}{#2\S#1#3}
\crefformat{subsubsection}{#2\S#1#3}
\crefformat{figure}{#2Fig. #1#3}
\crefformat{table}{#2Table #1#3}

\usepackage[labelformat=simple]{subcaption}
  
\usepackage{multirow}
\usepackage{siunitx}
\usepackage{makecell}
\usepackage{pgf}
\usepackage{pgfplots}
\pgfplotsset{compat=1.18}
\usepackage{url}
\setuptodonotes{inline}

\usepackage{tikz}   
\newcommand{\CRicon}[1]{
	\tikz[baseline=(char.base)]{
		\node[shape=circle, rounded corners=3pt, inner sep=0pt,
		top color=black, bottom color=black, text=white,
		font=\bfseries] (char)
		{#1};
	}
}

\usepackage{circledsteps}
\newcommand\myCircled[2][]{\ifmmode
\Circled[fill color=black,inner color=white,#1]{\mathsf{#2}}
\else
\Circled[fill color=black,inner color=white,#1]{\sffamily#2}
\fi
}

\newtcolorbox{insightbox}{
    colback=purple!10, colframe=purple!80,
    title=Insight:, fonttitle=\bfseries, sharp corners,
    top=2mm, bottom=2mm, left=1mm, right=1mm
}

\definecolor{asparagus}{rgb}{0.53, 0.66, 0.42}

\author{Yun-Chih Chen}
\author{Tristan Seidl}
\author{Nils Hölscher}
\author{Christian Hakert}
\author{Minh Duy Truong }
\author{Jian-Jia Chen}
\authornote{These authors contributed equally to this work.}
\affiliation{
  \institution{TU Dortmund University}
  \city{Dortmund}
  \country{Germany}
}
\email{{yunchih.chen, tristan.seidl, nils.hoelscher, minhduy.truong, jian-jia.chen}@tu-dortmund.de}

\author{João Paulo C. de Lima\textsuperscript{$\dagger$}}
\author{Asif Ali Khan}
\author{Jeronimo Castrillon\textsuperscript{$\dagger$}}
\authornotemark[1]
\affiliation{
  \institution{Technische Universität Dresden, ScaDS.AI\textsuperscript{$\dagger$}}
  \city{Dresden}
  \country{Germany}
}
\email{{joao.lima, asif_ali.khan, jeronimo.castrillon}@tu-dresden.de}

\author{Ali Nezhadi}
\author{Lokesh Siddhu}
\author{Hassan Nassar}
\author{Mahta Mayahinia}
\author{Mehdi Baradaran Tahoori}
\author{Jörg Henkel}
\authornotemark[1]
\affiliation{
  \institution{Karlsruhe Institute of Technology (KIT)}
  \city{Karlsruhe}
  \country{Germany}
}
\email{{ali.nezhadi, lokesh.siddhu, hassan.nassar, mahta.mayahinia, mehdi.tahoori,henkel}@kit.edu}

\author{Nils Wilbert}
\author{Stefan Wildermann}
\author{Jürgen Teich}
\authornotemark[1]
\affiliation{
  \institution{Friedrich-Alexander-Universität Erlangen-Nürnberg}
  \city{Erlangen}
  \country{Germany}
}
\email{{nils.wilbert, stefan.wildermann, juergen.teich}@fau.de}

\begin{document}

\title{Modeling and Simulating Emerging Memory Technologies: A Tutorial}
\maketitle
\renewcommand{\shortauthors}{Y. Chen et al.}
\authornote{These authors contributed equally to this work.}

\section*{Abstract}
Non-volatile Memory (NVM) technologies present a promising alternative to traditional volatile memories such as SRAM and DRAM. Due to the limited availability of real NVM devices, simulators play a crucial role in architectural exploration and hardware-software co-design. This tutorial presents a simulation toolchain through four detailed case studies, showcasing its applicability to various domains of system design, including hybrid main-memory and cache, compute-in-memory, and wear-leveling design. These case studies provide the reader with practical insights on customizing the toolchain for their specific research needs. The source code is open-sourced.

\section{Introduction}\label{sec:Introduction}
As semiconductor technology advances, energy efficiency is becoming a significant barrier to scalability.
This challenge has resulted in the development of heterogeneous computing devices and the \textit{Dark Silicon} phenomenon, in which portions of a chip are selectively turned off to save energy.
Non-volatile memory (NVM) technologies are a promising solution because they provide energy-efficient alternatives to traditional volatile memory.
Historically confined to storage applications such as SSDs and read-only instruction storage in embedded systems, new NVM technologies with access latencies approaching those of DRAM \cite{pcm:2024, 10145822} have opened the door for their use as main memory with a moderate performance penalty.

However, diverse application contexts expose different design requirements.
On the one hand, small and remote devices (e.g., edge servers at traffic lights or base stations in hard-to-access locations) may operate for extended durations with limited maintenance. In these settings, high power consumption, frequent dirty data writebacks, and costly data restoration upon wake-up are especially problematic for current memory technologies, including DRAM, SSD, and eFlash~\cite{8351201}.
On the other hand, data-center workloads continue to grow in scale and complexity, pushing SRAM and DRAM to their limits in terms of performance, bandwidth, density, and energy usage. 
This has driven the evolution of memory technology toward greater heterogeneity and specialization.
NVMs, with their longer retention times, are well-positioned to play an important role in this transition, particularly as an adaptable, energy-efficient memory option for a broader spectrum of applications, ranging from power-constrained wearable devices up to large-scale enterprise systems.

However, to fully realize NVMs' potential in mainstream computing, extensive research in computer architecture is required to address their challenges
, including limited endurance, higher fault rates, and expensive write operations.
These characteristics necessitate adaptive hardware-software co-design solutions to improve performance and longevity.
Effective co-design strategies must align software requirements with hardware limitations using techniques such as wear leveling, error correction, and vertical integration, ensuring that both software and hardware work together to maximize NVMs' capabilities.

Recognizing the importance of addressing these challenges as part of a broader effort to advance memory technologies, the German Research Foundation (DFG) launched SPP 2377 in 2022.
This 6-year national priority program brings together 23 principal investigators from 13 institutions and is subdivided into 14 research projects spanning topics such as computer architecture, databases, compilers, and operating systems.
The program investigates a wide range of issues related to emerging memory technologies, including commercially available ones, such as FRAM, Intel Optane DIMMs, and UPMEM\cite{devaux2019true}, as well as experimental NVMs that are still in development.

This paper is based on a 3-hour tutorial session delivered at the ESWEEK 2024 conference. 
It describes 
a highly configurable NVM simulation toolchain, developed in a collaborative effort within the SPP 2377. 
At its core, the toolchain builds atop gem5~\cite{binkert:2011} and NVMain (using NVMain2.0~\cite{poremba:2015}). 
With this, we support further advancements in the field by making our extensions accessible to the broader research community.

The toolchain enables the research community to do architectural exploration and study the behavior of target applications before the NVM devices become commercially available. 
This paper introduces the reader to the basic operation of the toolchain and provides a hands-on guide for the usage of our extensions, presented through a series of case studies that demonstrate the types of architectural exploration the toolchain enables.
We focus on four NVM research issues to illustrate the toolchain’s capabilities.
The first case study focuses on hybrid main memory (\cref{sec:nvm-dram}).
The second case study focuses on trace generation and trace-based wear-out analysis, which allows designers to examine memory access patterns and assess the behavior of target applications (\cref{subsec:tudo}).
The third case study explores architectural trade-offs in NVM-SRAM hybrid cache designs (\cref{sec:nvm-cache})
Finally, the fourth case study demonstrates the use of NVM in Compute-in-Memory (CiM) architectures (\cref{sec:nvm-cim}). 

\section{Background}\label{sec:Background}

NVM allows devices to wake up, operate, and power down without transferring data between memory and storage. NVM also significantly reduces standby power consumption compared to SRAM.  Lastly, most NVMs allow storing multiple bits per memory cell, thus increasing memory density.
From wearables and implantables to IoT and edge devices, NVM has the potential to enable persistent low power operation.
For example, a cell phone could extend its battery life by incorporating a low-power NVM-based co-processor alongside its high-performance processor.
The co-processor could monitor inputs, such as audio commands, while the phone is in sleep mode, waking the main processor only when necessary.

NVM can drive further advancements in scaling advanced data center SoCs, such as the ARM Neoverse series~\cite{ArmN1}, which feature hundreds of cores per chip. These SoCs operate at the edge of physical limitations, making power efficiency and effective cooling critical to maintaining performance without thermal throttling.  Research on advanced sub-1nm process nodes reveals that, depending on operating temperature, SRAM-based L2 caches can make up 25\% to 50\% of data center-grade SoCs' overall power consumption \cite{thermal_soc:24}.  On the other hand, NVM offers an energy-efficient, area-efficient alternative to SRAM, with the potential to significantly reduce the leakage current \cite{10750212}.

\subsection{NVM Technologies}
\label{subsec:materials}
There are many different types of NVM technologies.
Some rather mature NVM technologies include: Phase Change Memory (PCM), Resistive Random-Access Memory (ReRAM), Ferroelectric RAM (FRAM), and Spin-Transfer Torque RAM (STT-RAM).
Unlike volatile memory, which stores data using electric charge, many NVM technologies represent data through resistance levels, but each technology uses different materials and distinct physical principles to alter the state of the memory cell. 
PCM uses heat to change the material state between crystalline and amorphous, while ReRAM moves ions within a dielectric material. On the other hand, STT-RAM encodes data through the magnetic orientation of magnetic tunnel junctions (MTJs).

When PCM was introduced to the market as Intel Optane Persistent Memory in 2019, it was unable to compete with DRAM due to cost concerns. However, recent developments in material science research have revived interest in its potential, especially for processing-in-memory architectures, because of its exceptional endurance (i.e., $2 \cdot 10^8$ cycles) and quick access times (i.e., 40 ns) \cite{pcm:2024}.
ReRAM as a replacement for SRAM can address the power and cooling issue for ultra-dense 3D ICs with tightly coupled logic and memory \cite{rram:23}. Notably, recent fabrication of ReRAM-based AI accelerators has shown significant energy savings by eliminating the need to load weights from off-chip memory \cite{rram:22}.
FRAM, with its fast read/write speeds of around 50\,ns, low power consumption, and high switching durability (i.e., $10^{13}$ cycles), has been widely adopted in RFID tags and microcontrollers \cite{fram:24}.

For a detailed comparison of NVM material and architectural exploration for trade-offs in access latency, area, power efficiency, and endurance, we advise taking a look into Pentecost~et~al.'s comprehensive survey \cite{9773239}.

\subsection{Commercial Maturity of NVM}
\label{sec:commercial_maturity}
Commercially, NVMs have 
been in mass production for years. 
For example, Samsung began STT-MRAM's mass production in 2019.  8Mb of this NVM is used in Sony’s GPS receiver chip as part of Huawei GT2 smartwatch \cite{10145822}.  The NVM enables a remarkable two-week battery life, far exceeding the few days of a volatile-memory-based smartwatch. In the data center market, Intel’s release of Optane Persistent Memory (Optane DCPMM) in 2019 brought significant attention to in-memory data structures for persistent memory. However, the discontinuation of Optane in 2022 due to low profitability reflects a broader issue: NVM’s pricing remains uncompetitive with mature memory technologies. As shown in \cref{table:price_per_bit}, data from DigiKey as of November 2024 \cite{digikey} reveals that NVM capacities are still significantly more expensive than DRAM or NAND flash. While this cost disparity discouraged some researchers from exploring NVM further, NVM is finding success in specialized markets where its unique capabilities align closely with application needs.

For example, NXP plans to integrate NVM into automotive microcontrollers, while Samsung and Sony are utilizing NVM as a frame-buffer memory in advanced image sensors. Avalanche Technology has proposed deploying an 8 Gb MRAM package in low earth orbit for space applications. In these specialized domains with well-defined software behavior, NVM provides an opportunity for tight hardware-software co-design. Unlike general-purpose computing, where hardware is designed to accommodate any software, these domains enable tailoring of NVM’s capabilities to specific applications.

While NVM alone might be expensive, pairing NVM with mature memory technologies proves cost-effective. IBM, e.g., integrated Everspin’s 1 Gb STT-MRAM chips into the IBM FlashCore Module, an enterprise-grade solid-state drive, to achieve high reliability.  Another promising use case is an SRAM and STT-MRAM hybrid last-level cache, as we explore in \cref{sec:nvm-cache}.
The low profitability of NVM also stems from misconceptions about its role. Treating NVM as a drop-in replacement for DRAM imposes stringent requirements—such as long retention times, errorless access, and low latency—that drive up costs unnecessarily. Many applications do not require these demanding features, but aligning application requirements with NVM capabilities requires tight hardware-software co-design. This process, however, demands extensive testing and iteration. Simulators play a critical role in this, offering a cost-effective and flexible way to explore hardware-software co-design strategies, optimize NVM integration, and adapt its capabilities to meet application-specific needs.

\begin{table}[h!]
    \centering
    \begin{tabular}{|l|c|c|c|}
        \hline
        \textbf{Memory Type} & \textbf{Price (USD)} & \textbf{Capacity} & \textbf{Price per gigabit (USD)} \\ \hline
        Infineon FRAM        & 50                  & 16 Mbit           & $3125 $ \\ \hline
        Infineon ReRAM        & 10                  & 512 Mbit          & $19.53$ \\ \hline
        Everspin STT-MRAM    & 100                 & 1 Gbit            & $100$ \\ \hline
        Micron DRAM          & 108                 & 256 Gbit          & $0.42 $ \\ \hline
        Kioxia NAND Flash    & 132                 & 8 Tbit            & $0.0165$ \\ \hline
    \end{tabular}
    \caption{Price, capacity, and price per bit for various memory types (source: Digikey).}
    \label{table:price_per_bit}
\end{table}

\subsection{NVM Simulation}
\label{subsec:nvm-simulation}
To effectively adapt application-specific needs to a target NVM material’s unique characteristics (e.g., endurance limitations and variable access latencies), researchers must iteratively explore co-design strategies that align hardware capabilities with software requirements.  
A significant barrier to advancing NVM research is the lack of accessible experimentation platforms. Academic researchers face the difficulties of limited access to real NVM devices, which are often proprietary, while memory manufacturers encounter high costs associated with evaluating software behavior on physical NVM chips, given their relatively slow access times and destructive write operations.

To overcome these limitations, software-based simulations offer a cost-effective means of exploring the design space. They enable detailed modeling of system components and interactions, facilitating iterative co-design between hardware and software. Simulations allow researchers to test architectural parameters, analyze memory behavior, and develop optimization techniques without the constraints of physical hardware.
An example, among many, is recent work that investigates the effect of replacement policies on the lifetime of NVM caches~\cite{escuin_hpca23}. 

There are two primary simulation methods widely used in NVM research:
(a) \textit{Cycle-accurate full-system simulations} provide precise modeling of all system components and produce results that closely approximate running on real device behavior. While this precision captures intricate interactions such as cache and TLB behaviors, it comes at the cost of significantly longer execution times.
(b) \textit{Trace-based simulations} run on pre-produced logs of memory accesses.  It offers a faster alternative to cycle-accurate approaches, making them suitable for tasks like analyzing memory usage patterns or evaluating wear-leveling algorithms. However, their accuracy depends heavily on the quality and comprehensiveness of the traces used.

\subsection{Studied Toolchains}
\label{subsec:toolchain}
\begin{figure}[t]
    \includegraphics[width=0.8\columnwidth]{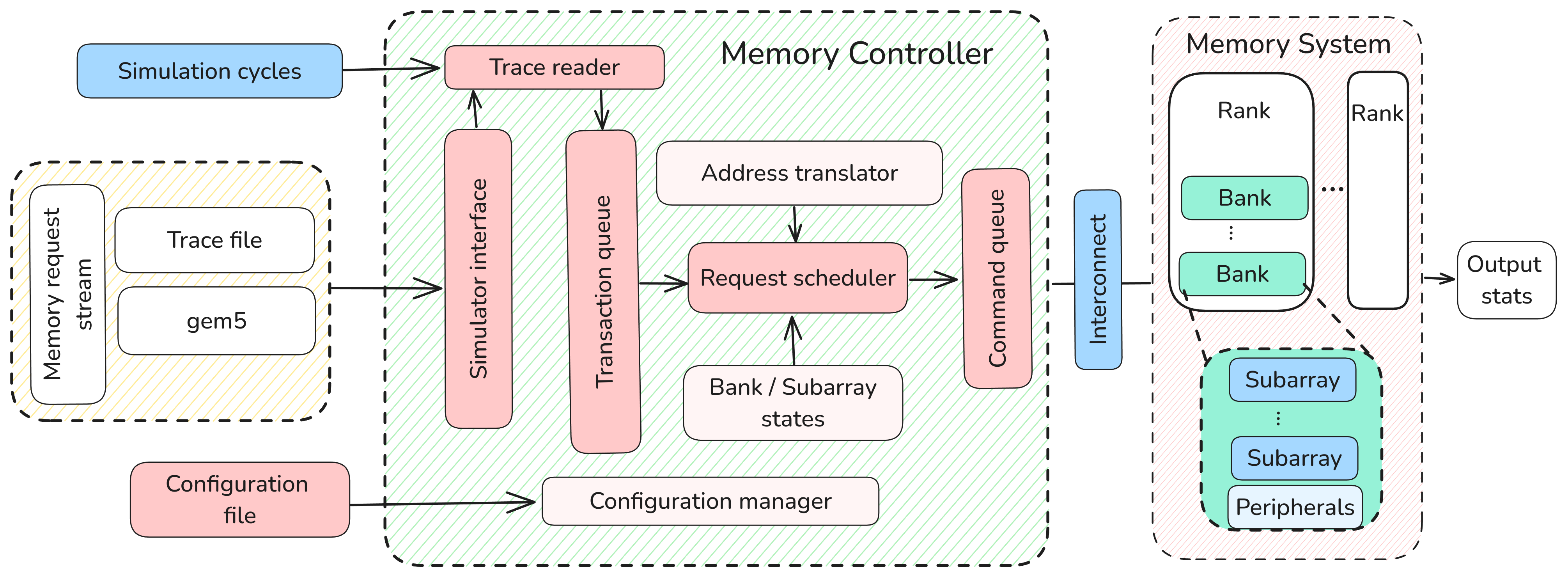}
    \caption{
        Overview of the NVMain flow. Adapted from~\cite{poremba:2015,khan2019rtsim}.
    }
    \label{fig:nvmain}
\end{figure}

In this work, we focus on cycle-accurate full-system simulation for NVM research and offer a tutorial on a toolchain that consists of gem5 and NVMain. The toolchain is widely used in the research domain of emerging memory technologies~\cite{dwm-nvmain:19, hakert:2020:base, hakert:2022, hoelscher:2022, khan2019rtsim, khan2023downshift, asif:18, Wilbert:2024a, Wilbert:24b}. 
The tutorial is composed of four case studies that showcase our extensions to the toolchain, aiming to familiarize the reader with a powerful utensil for their NVM research.

gem5~\cite{binkert:2011} is a widely used cycle-accurate full-system simulator for computer architecture research. It models critical side effects, including cache set behavior, TLB misses and hits, and internal memory device states such as buffers and row hits.
NVMain~\cite{poremba:2015} extends gem5 to model NVM technologies alongside DRAM and SRAM. \cref{fig:nvmain}
shows an overview of NVMain.
It can be configured to emulate the timing and energy parameters corresponding to different NVM technologies, as well as to support detailed investigations of memory behavior.
Its subarray modeling captures core memory operations, such as the varied access latency of multi-level cell (MLC).  It also performs access accounting, which reflects memory wearing and enables studies of endurance enhancement.
NVMain also offers fault injection to analyze the impact of memory defects and failures.

To ease evaluation, the toolchain comes with a set of benchmarks that offer diverse memory access patterns. This set was assembled throughout our previous work~\cite{hakert:2020:base, hakert:2020:softwear, hakert:2022, hoelscher:2023} and it includes benchmarks from MiBench~\cite{guthaus:2001} as well as from Olden~\cite{rogers:1995}. These benchmarks were previously~\cite{hoelscher:2023} ported to Unikraft~\cite{kuenzer:2021}, which we added to the toolchain in~\cite{hakert:2020:split}. Unikraft is a unikernel kit that bundles the required kernel functionality with the application into a single binary. This minimizes binary size by excluding unnecessary services and consolidates all functionality into a single address space, where the latter can simplify memory access analysis. Unikraft is highly compatible with Linux, which makes it simple to port Linux applications to Unikraft.

\section{Case Studies}
The four case studies presented in this tutorial paper are as follows: 

\begin{enumerate}
    \item \textbf{Memory Subsystem Modeling:} 
     This case study introduces the configuration (e.g., timing and scheduling) and simulation of DRAM and NVM-based main memories, and the implementation of custom memory operations in NVMain, such as RowClone for in-memory bulk copying.  By the end of this case study, readers will be equipped with the knowledge to evaluate and refine memory system designs, enabling them to make informed decisions tailored to specific application needs.

    \item \textbf{Trace Writer:}  This case study introduces the trace-writing capabilities of NVMain2.0, providing step-by-step guidance on integrating and customizing trace writers for diverse use cases. The reader will learn how to set up a basic trace writer, integrate it into the toolchain’s build process, and log essential memory access information. Additionally, this case study explores how to analyze logs for tracking NVM wear-out and examining memory access patterns. By the end, the reader will be equipped with the knowledge to extend NVMain2.0’s trace-writing functionality for specific research or development needs.
 
    \item \textbf{NVM-SRAM Hybrid Caches:} This case study presents an architectural exploration of hybrid caches consisting of SRAM and STT-RAM cells. The reader will learn how to configure asymmetric read/write latencies for different memory materials, how to vary the non-volatility ratio, and how to evaluate their energy and performance impact, demonstrated through an image processing task (read-intensive) and a merge sort algorithm (write-intensive). The case study enables exploration of novel cache replacement policies and evaluating the trade-offs of hybrid memory systems specific to applications of interest.
    
    \item \textbf{Compute-in-Memory (CiM):} 
    This case study details the step-by-step modifications required to integrate Compute-in-Memory (CiM) functionality into gem5, including extending memory controllers, implementing CiM-specific operations, and adapting simulation parameters for accurate modeling.
     The resulting CiM-enabled system is evaluated with several real-world applications.
    The reader gains insights into the architectural differences between CiM and traditional Processing-in-Memory (PIM), the trade-offs of different CPU-CiM communication methods, and the physical placement of CiM components within memory modules.

\end{enumerate}

For setting up the toolchain, please see the setup instructions provided in appendix~\cref{app:setup}.

\subsection{DRAM and NVMs-Based Main Memory Simulation}
\label{sec:nvm-dram}

\subsubsection{Overview}
As stated in Section~\ref{sec:Introduction}, NVMs have the potential to replace conventional SRAM and DRAM technologies but require careful design decisions to mitigate the impact of their challenges, e.g., non-reliability, expensive write operations, on the running application's accuracy, overall performance and energy consumption. Various optimizations have been proposed to achieve this including: novel architectural designs, memory controllers, and efficient scheduling strategies~\cite{5416645, 1555815}.

Nonetheless, choosing the right memory technology and determining the best set of optimizations for a specific application domain is non-trivial and requires system-level simulations to validate and refine design and optimization decisions. To this end, this case study provides a step-by-step introduction to the memory subsystem design, focusing on understanding its key components and comparing DRAM and NVM technologies using NVMain. As explained in Section~\ref{subsec:toolchain}, the modular structure and various simulation modes of NVMain enable quick prototyping and evaluation of memory systems, whether for specific application needs or to enhance overall system performance.

\subsubsection{Objective}
\label{subsubsec:tudr:objective}
This case study aims to familiarize readers with memory subsystem simulation by guiding them through the various components of NVMain. Specifically, it introduces readers to exploring architectural features, reordering memory requests to improve row buffer hits, and leveraging write and read queues to optimize NVMs' access latency. Since we cannot cover all parts of the memory subsystem, we intentionally select different experiments related to memory system modeling and optimizations, providing insights that enable readers to make tailored modifications of their choice. The case study also includes an introduction to configuring and simulating hybrid main memory systems, i.e., combining DRAM and NVM technologies. Finally, we show how to implement new memory operations exemplified by RowClone, a technique that enhances the functionality of modern memory systems for efficient bulk copies in memory.

\subsubsection{Methodology}
\label{subsec:nvm-dram-methods}
Before delving into the different experiments,  let us first introduce key simulation parameters, including memory timings and scheduling policies.

~\\\noindent \textbf{Timing in DRAM and Non-Volatile Memories:}\\
\textit{Single-Bank Timing:}
In DRAM, a read cycle requires restoring data (tRAS = activation + restoration time) and starting precharge before closing a row, as reads destroy data in the capacitor. The read cycle time (tRC) includes activation (tRCD), restoration, and precharge (tRP). In contrast, read operation in NVMs is non-destructive, so restoration is not needed. The read cycle for NVMs is dominated by tRCD + tBURST. 
\textit{Inter-Bank Timing:} To manage power and limit peak current, tFAW (four activation window) restricts four activations within a set time, while tRRD (Row-to-Row Activation Delay) spaces out activations to prevent excessive current. These rules apply within and across bank groups. In NVMs, high write currents and long write times can overlap between writes and activations, managed using parameters like tWWD (Write-to-Write Delay), tWAD (Write-to-Activation Delay), and tAWD (Activation-to-Write Delay).

~\\\noindent \textbf{Scheduling Policies:}\\
Memory controller's scheduling policies significantly affect the overall performance. Below are some commonly used policies implemented in NVMain:\\
\textit{First-Come-First-Served (FCFS):}
In single-bank timing, requests are processed in their arrival order without optimizing for delays caused by row activation or precharge. This hurts inter-bank parallelism, as idle banks must wait for pending requests in busy banks.\\
\textit{First-Ready First-Come-First-Served (FR-FCFS):}
In single-bank timing, FR-FCFS reorders requests to maximize row-buffer hits; leading to a reduced number of row activations and precharges that reduce the access latency. For inter-bank timing, parallelism is maximized by overlapping requests across banks, enabling some banks to process read/write requests while others handle activations or precharges.\\
\textit{First-Ready First-Come-First-Served with Write Queue Flush (FR-FCFS-WQF):}
In single-bank timing, writes are batched when the write queue fills, minimizing write-to-read penalties. For inter-bank timing, write batching is coordinated across banks to prevent stalling reads, improving efficiency and reducing delays.

In the following, we outline instructions to guide readers to conduct experiments and achieve the desired objective (see Section~\cref{subsubsec:tudr:objective}). The are total four tasks in this case study that are organized into two categories: (i) full-system simulation using gem5 and NVMain, (ii) Stand-alone trace-driving simulation using NVMain. The first two tasks focus on exploring the performance trade-offs of memory controller scheduling policies and hybrid memory setups, and the last two tasks cover trace-based simulations and the implementation of new operations in NVMain. All file paths used in these instructions as well as the current working directory are relative to \texttt{root/simulator/}, with root being the toolchain's repository's root directory. 

~\\\noindent \textbf{Performance Comparison of Scheduling Policies:}\\
The impact of different scheduling policies on performance is illustrated by compiling the matrix multiplication (MM) application (see Appendix~\ref{app:tudr:scheduling}) and executing it on gem5 with a desired configuration. The scheduling policy (FCFS, FRFCFS, or FRFCFS-WQF) can be configured in the configuration file, \texttt{nvmain/Config/PCM\_ISSCC\_2012\_4GB.config}, before running any simulation. For instance, changing the memory controller policy from the \texttt{FCFS} to \texttt{FRFCFS} or \texttt{FRFCFS-WQF} and examining statistics such as the number of row buffer hits and misses and the access latencies (\texttt{rb\_hits, rb\_miss, averageLatency and averageTotalLatency}), provides insight into performance differences.
\texttt{averageLatency} reflects the memory module's internal service time, excluding system-level queuing delays, while \texttt{averageTotalLatency} represents end-to-end latency.  
For \texttt{FCFS}, \texttt{rb\_hit} and \texttt{rb\_miss} are both zero, with \texttt{averageTotalLatency} in the millisecond range.  
With \texttt{FRFCFS}, row buffer hits appear but remain low and \texttt{averageTotalLatency} is reduced to $\sim6\mu s$.  
\texttt{FRFCFS-WQF} introduces separate read/write queues, significantly increasing \texttt{rb\_hits} beyond misses. While queue latency rises, \texttt{averageLatency} improves, indicating less read/write operations on NVM cells.

~\\\noindent \textbf{Hybrid Memory Simulation:}\\
Hybrid main memories that combine DRAM and NVM have been widely explored in research and industry. gem5 supports the simulation and evaluation of configurations integrating DRAM and NVM as main memory. These configurations leverage NVM's high density and persistence alongside DRAM's speed, enabling the design of architectures that enhance memory capacity and efficiency while addressing data retention challenges.

The \texttt{hybrid\_example.py} file provides an example configuration of such a hybrid memory system.
To test this configuration, Appendix~\cref{app:tudr:hybrid} presents some commands to run the STREAM benchmark~\cite{stream_bench}.
In the \texttt{stream.c} file, lines \texttt{181–196} demonstrate how to allocate arrays in the data segment as pinned memory statically. The memory address ranges are defined in line \texttt{89} of \texttt{hybrid\_example.py}.
Again, we refer to the commands in Appendix~\cref{app:tudr:hybrid} to compile and simulate the stream kernel with the desired pinned memory size.

The performance results of the hybrid memory configuration on the STREAM benchmark demonstrate improvements in both latency and bandwidth compared to a DRAM-only system. The hybrid setup balances the strengths of DRAM’s high-speed access with the cost-effectiveness and persistence of NVM. 
Further system-level optimizations, such as more advanced memory scheduling policies and fine-tuned data placement strategies, could further leverage the potential of hybrid memory systems for diverse application scenarios.
Section~\ref{sec:nvm-cache} provides another use case of hybrid memory simulation for hybrid caches.

\begin{figure}[ht!]
    \centering
    
    \begin{subfigure}[b]{0.45\textwidth}
        \centering
        \includegraphics[width=\textwidth]{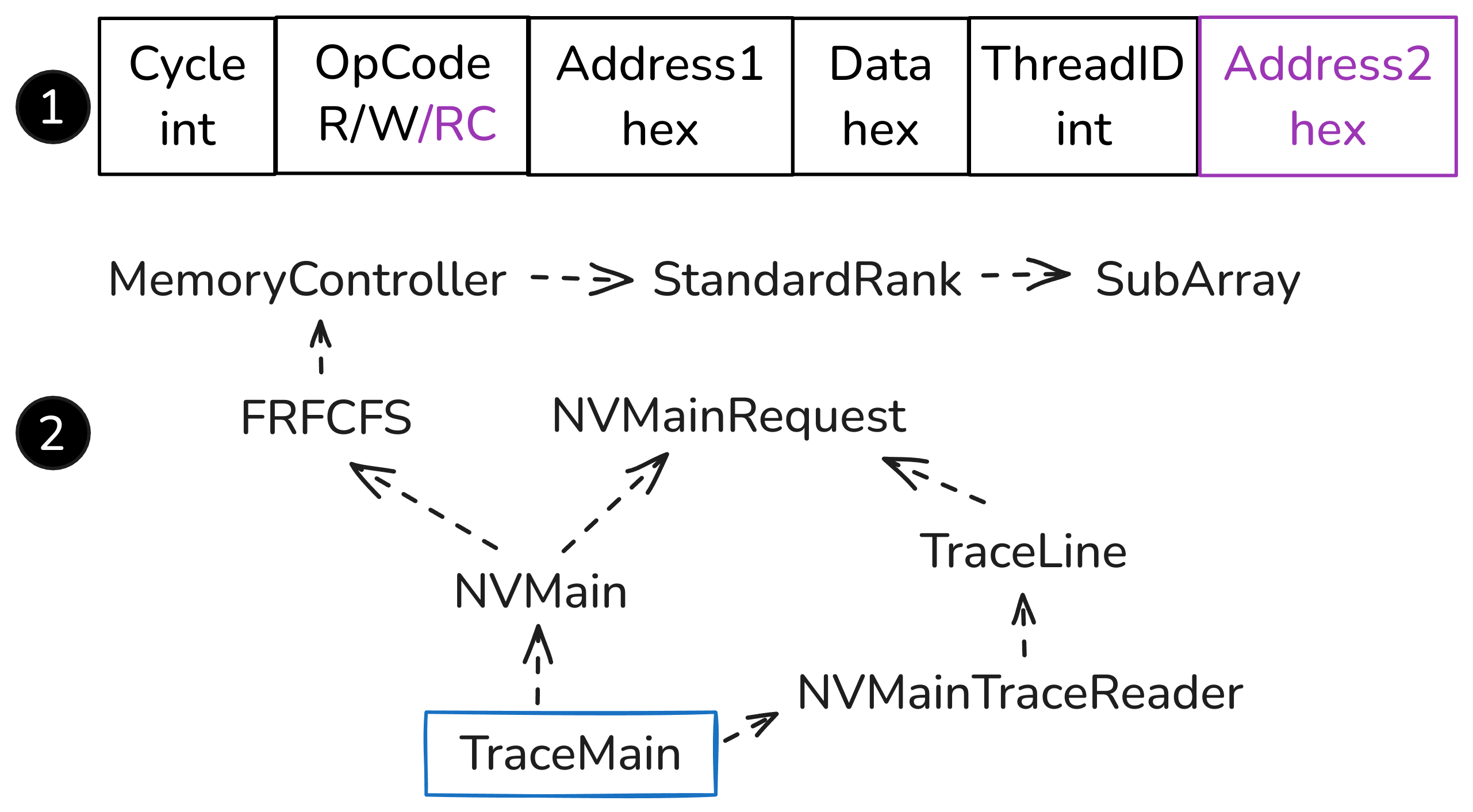} 
        \caption{NVMain classes to modify}
        \label{fig:nvmain_changes}
    \end{subfigure}
    \hfill
    
    \begin{subfigure}[b]{0.45\textwidth}
        \centering
\begin{tabular}{clll}
\hline
\multicolumn{1}{l}{}                                                                                             & \textbf{Copy}   & \textbf{Zero} &        \\ \hline
\multirow{3}{*}{\begin{tabular}[c]{@{}c@{}}Memory\\ Energy\\ (µJ)\end{tabular}} & Baseline                         & 3.6        & 2.0  \\
                                                                                & FPM~\cite{ambit}                  & 0.04        & 0.04 \\
                                                                                & FPM (NVMain)                     &  0.04       &   0.05\\ \hline
                                                                                
\multirow{3}{*}{\begin{tabular}[c]{@{}c@{}}Latency\\ (ns)\end{tabular}}         & Baseline                         & 1046        & 546 \\
                                                                                & FPM~\cite{ambit}                  & 73        & 73  \\
                                                                                & FPM (NVMain)                     &  90       &  90 \\ \hline
\end{tabular}
        \caption{Latency and energy validation}
        \label{fig:rc_validation}
    \end{subfigure}

    \caption{Overview of the modifications made to NVMain to support RowClone and its validation}
    \label{fig:combined}
\end{figure}

~\\\noindent \textbf{Trace-Driven Simulation:}\\
NVMain also supports trace input files (see Figure~\ref{fig:nvmain}), which must follow the specified format (\myCircled{1}) in Figure~\ref{fig:nvmain_changes}.
To simulate a trace file on NVMain (either in \texttt{fast}, \texttt{debug}, or \texttt{prof} mode), the configuration file, the trace file, and the number of simulation cycles should be specified.
Appendix~\cref{app:tudr:traceSim} illustrates the specific commands to view an example trace file and run it using NVMain. The trace-driven simulation specifically focuses on the memory subsystem (and not on the full system) and enables using NVMain with other simulators/frameworks, i.e., the input trace does not necessarily need to come from gem5. It also provides more flexibility and control over the simulation which is why we will use it in the next task to introduce new memory operations. 
The integrated full-system simulation with gem5 would require extending a few classes in the memory model of gem5 to make it work. For instance, if the new operation is neither a read nor a write, a new type should be specified; if the operation has to be exposed to the programmer, a new instruction or pseudo-instruction should be introduced.
Additionally, changes would be required to the cache coherence protocol to ensure data coherence in CPU caches. 

~\\\noindent \textbf{Adding a New Operation:}\\
NVMain can be extended to support new operations and simulate tailored memory system designs, e.g., for CiM (Compute-in-Memory). For instance, it has been demonstrated by previous research that bulk data copying and initialization can be done within the main memory to eliminate data transfer over the external bus.
For DRAM, the RowClone \cite{kit/rowclone} mechanism can copy an entire row from a source address to a destination address via back-to-back \emph{activation}. Similarly, in-memory mirroring \cite{rowMirroring} performs copy directly within the resistive crossbar memories without the need to read the data.

To implement RowClone in NVMain, multiple classes need to be modified, as summarized in \myCircled{2} of Figure~\ref{fig:nvmain_changes}. 
Copying a source row (\texttt{src}) to a destination row (\texttt{dst}) falls into one of these two cases: both rows are within the same subarray, in which case the RowClone (\texttt{RC}) is performed using Fast Parallel Mode (FPM); or \texttt{src} and \texttt{dst} reside in different banks or different subarrays within the same bank, hence accomplished by the Pipelined Serial Mode (PSM)~\cite{kit/rowclone}. 
In the provided code (details in Appendix~\ref{app:tudr:adding_ops}), the RowClone functionality is limited to FPM. The code parses memory commands whose opcode is \texttt{RC} and contains two addresses, as highlighted in \myCircled{1}. 
To support FPM, the subarray model accepts back-to-back \texttt{ACTIVATE}s if they belong to the same subarray, otherwise, it drops the second \texttt{ACTIVATE}.
Appendix~\ref{app:tudr:adding_ops} presents commands to compile NVMain as a stand-alone tool and simulate \texttt{RC} operations from a trace file.
Table~\ref{fig:rc_validation} presents the memory latency and energy improvements achieved by FPM RowClone for bulk copying and zeroing 4KB of data. The baseline for comparison is the same operation performed by the CPU (details in Appendix~\ref{app:tudr:adding_ops}). FPM~\cite{ambit} refers to the estimations provided in the original paper, while FPM (NVmain) refers to our validation results from this case study. 

\begin{insightbox}

This case study provided insights into customizing NVMain, both stand-alone and with gem5, covering main memory simulations based on DRAM, NVM, and hybrid DRAM-NVM configurations, and new memory operations. Specifically, the RowClone operation was demonstrated as an example, highlighting its potential for efficient in-memory bulk copies.
\end{insightbox}
\subsection{Access Tracing and Data Processing Using Trace Writers}
\label{subsec:tudo}

\subsubsection{Overview}
\label{tudo:overview}
Emerging NVM technologies have the potential to complement or partially replace DRAM as main memory, but they face the challenge of wear-out over time.
If memory cell updates are unevenly distributed, heavily updated memory bits can wear out prematurely, reducing the overall capacity and lifespan of the memory. In the worst case, the memory will fail when the first cell is fully worn out. To mitigate this issue, wear-leveling techniques are essential.
There exists a wide variety of various wear-leveling approaches on hard- and software-level~\cite{yang:2007, PageWL, CacheWL, Loop2RecWL, hoelscher:2022, HeapStackWL, Stack, HeapWL}.
The data-comparison write scheme~\cite{yang:2007} is a hardware-based technique that reduces unnecessary write operations. The memory cell is only updated when the target value differs from the one currently stored. For most NVMs, wear is only induced by writing but for some technologies, such as FeRAM~\cite{philofsky:1996}, it can also be inflicted when reading.
Software-based wear-leveling solutions can be better tailored to the application's needs by focusing on important memory regions or reacting to specific access patterns. Usually, such solutions involve more spatial and computational overhead compared to hardware-based solutions but they might be applied when a suitable hardware-solution does not exist.

While many wear-leveling solutions have gained popularity, their evaluations often lack depth. In our previous work \cite{hoelscher:2023}, we built on the toolchain presented in this paper to develop a new wear-analysis method that provides a detailed assessment of NVM wear-out at a single-bit granularity.
This approach tracks the number of bit flips per memory cell during an application’s simulation, as bit flips are directly proportional to wear. The collected data is then recorded in a trace for further analysis, applying metrics that provide insights into how applications induce different degree of wear on NVM. 
This detailed wear analysis helps refine and optimize wear-leveling strategies while making previously in-house evaluation methods more comparable.

Besides  wear-out analyses, the simulation toolchain described in this study can provide insights into many aspects of memory behavior. For example, it may monitor library usage patterns \cite{hakert:2020:split} and examine read access patterns \cite{hakert:2022}.  This may be used to assess cache replacement policies, debug and profile workloads, and discover abnormal memory access patterns or security risks such as side-channel attacks.
In summary, the cycle-accurate full-system simulation provides information beyond wear-out patterns, making the trace writer a very useful tool for broader system research.

\subsubsection{Objective}
\label{tudo:objective}
In this case study, we want to familiarize the reader with NVMain2.0's trace writing capabilities, enabling them to utilize these to their full extent for various use cases. To this end, we provide step-by-step instructions that guide the reader through the creation of a new trace writer. We cover a) the creation of a basic trace writer skeleton and its integration into the toolchain's build process, b) basic logging functionality and available simulation information, and c) processing during tracing for additional information and further analysis.

\subsubsection{Step-By-Step Instructions}
\label{subsubsec:tud-instructions}
In this section, we guide the reader iteratively through the creation of a new trace writer in NVMain2.0, such that they can set up a suitable instance for their respective use case. Given the root directory \texttt{NVM\_Simulation} of the toolchain's repository, all paths mentioned in this section are relative to \texttt{NVM\_Simulation/simulator/nvmain/traceWriter/}.

~\noindent \textbf{Skeleton and Build Integration:}\\
For any use case, a trace writer's general setup remains the same. First, we show how to create a basic trace writer skeleton and integrate it into the toolchain's build process. Start by creating a class for the \emph{TutorialTraceWriter} (\texttt{*.h/*.cpp)} that inherits from the base class \texttt{NVM::GenericTraceWriter}. The base class is provided by NVMain2.0 for custom trace writers to ensure an interaction with the simulation is possible, i.e., known hook-functions that can be called regularly by the simulation are present. Declare the functions shown in \cref{lst:skeleton} in the \emph{TutorialTraceWriter}'s header file.
\begin{lstlisting}[caption={Trace Writer Skeleton},label=lst:skeleton,language=c++,basicstyle=\ttfamily\scriptsize]
class TutorialTraceWriter : public NVM::GenericTraceWriter {
    public:
        virtual ~TutorialTraceWriter() = default;
        virtual bool SetNextAccess(NVM::TraceLine *nextAccess) override;
        virtual std::string GetTraceFile() override;
        virtual void SetTraceFile(std::string file) override;

    private:
        std::ofstream traceFile;
        std::string traceFileAddress;
    };
\end{lstlisting}
Additionally, include the variables shown in \cref{lst:skeleton}. They will store the resulting trace file's address as well as provide an output-file-stream required for the actual logging. The inherited function \texttt{GetTraceFile()} and \texttt{SetTraceFile()} are mandatory for NVMain2.0 to create the trace file in the first place. Define them in the trace writer's source file as shown in \cref{lst:getset}.
\begin{lstlisting}[caption={Getter/Setter Definition},label=lst:getset,language=c++,basicstyle=\ttfamily\scriptsize]
std::string TutorialTraceWriter::GetTraceFile() {
    return traceFileAddress;
}

void TutorialTraceWriter::SetTraceFile(std::string file) {
    traceFileAddress = file;
    traceFile.open(traceFileAddress.c_str());

    if (!traceFile.is_open()) {
        std::cout << "File could not be opened!" << std::endl;
    }
}
\end{lstlisting}

With the fundamental functions in place, the trace writer needs to be integrated into NVMain2.0's build process. For this, add all source files to the \texttt{SConscript}. It is part of SCons~\cite{scons}, the build system at hand, which lists all source files to consider when building. Furthermore, since NVMain2.0 uses the factory pattern~\cite{ellis:2007} to handle client code, the \emph{TutorialTraceWriter} needs to be added to the \texttt{TraceWriterFactory}'s source file. For both, stick to the practice of the current implementation in place. Finally, the system's configuration needs to be adapted. Open the NVM's configuration file that is to be used for your simulation, i.e., any \texttt{.../Config/*.config}. It contains the key parameters that specify the NVM's behavior. However, it is also used to configure the trace writer that will be used during the application's simulation, where three entries need to be modified. These are \texttt{PrintPreTrace}, allowing to toggle the trace writer's usage, \texttt{PreTraceFile}, specifying the resulting trace file's name, and \texttt{PreTraceWriter}, which indicates the trace writer that is to be used. Toggle the usage to \texttt{true}, set a trace file name and select the \emph{TutorialTraceWriter}. This concludes the skeleton's setup, which now can be built upon with meaningful logging behavior.

~\noindent \textbf{Logging Functionality:}\\
To enable logging, the trace writer provides an output-file-stream through which content can be put to the resulting trace. In context of these step-by-step instructions, the variable that stores the stream is \texttt{traceFile}, which you have added when setting up the trace writer skeleton. When the simulation is started, the output-file-stream will be opened and it can be written to with standard C++ I/O-stream operations. To get started with logging, go ahead and use the trace writer's destructor to put "Hello World" to the resulting trace. See \cref{lst:helloworld} for comparison. Build the toolchain, which is required because the modified trace writer needs to be compiled, and run the simulation as described in appendix~\cref{app:setup} using the \emph{helloworld}-benchmark. Check if you got the desired results.
\begin{lstlisting}[caption={Hello World},label=lst:helloworld,language=c++,basicstyle=\ttfamily\scriptsize]
TutorialTraceWriter::~TutorialTraceWriter() {
    traceFile << "Hello World" << std::endl;
}
\end{lstlisting}

Knowing how to add to the resulting trace, we move to modify the \emph{TutorialTraceWriter} such that it logs meaningful information on the simulated application. When simulating, the function \texttt{SetNextAccesss()} is called with every access to the NVM. All information on the access are passed to the function in form of \texttt{TraceLine} objects. By default, these objects contain information on:
\begin{itemize}
    \item The targeted physical address of memory the memory access (\texttt{GetAddress().GetPhysicalAddress()})
    \item The type of memory access, e.g., read or write (\texttt{GetOperation()})
    \item The data that will be put to / read from the memory address (\texttt{GetData()})
    \item The current cycle (\texttt{GetCycle()})
    \item The thread identification (\texttt{GetThreadId()})
\end{itemize}
In our previous work~\cite{hakert:2020:split}, we have modified the toolchain to additionally provide information on the current program counter. Knowing the application's memory layout, i.e., where the individual libraries are placed in the \texttt{.text}-section (see \cite{hakert:2020:split} for details), it allows to identify which part of the application has caused the respective memory access. You can access this information with \texttt{get\_program\_counter()}\footnote{
Requires to enable the bit flips simulation extension. See appendix~\cref{app:setup} for details.
}.

To complete this step of the instructions, go ahead and put some information available on the memory access to the resulting trace. We propose to log the access' targeted physical memory address, its operation and the last byte of the data that is to be read or written. Stick with the format \texttt{"Address | Operation | Data"}. Doing so should modify the function \texttt{SetNextAccess()} to look similar to \cref{lst:basiclog}~\footnote{
The function needs to return \texttt{true} to indicate that the memory access was handled correctly.
}

\begin{lstlisting}[caption={Basic TraceLine Logging},label=lst:basiclog,language=c++,basicstyle=\ttfamily\scriptsize]
bool TutorialTraceWriter::SetNextAccess(NVM::TraceLine *nextAccess) {
    const OpType operation = nextAccess->GetOperation();
    std::string opString;

    if(operation == NVM::READ) {
        opString = "READ";
    } else if(operation == NVM::WRITE) {
        opString = "WRITE";
    } else {
        opString = "OTHER";
    }

    traceFile << std::hex << nextAccess->GetAddress().GetPhysicalAddress() << std::dec << " | ";
    traceFile << opString << " | ";
    traceFile << static_cast<uint64_t>(nextAccess()->GetData().GetByte(0));

    return true;
}
\end{lstlisting}

~\noindent \textbf{Processing During Tracing:}\\
Simply logging available information, as shown previously, can be useful, but for some use cases, it may not be sufficient. In addition to storing data placement in the resulting trace, the trace writer can process memory access information to prepare the trace for further analysis or generate additional insights.
For example, in our previous work \cite{hoelscher:2023}, we compared the old content of accessed memory addresses to their new values and recorded the number of bit flips per memory cell. At the end of the simulation, the recorded data was stored in the trace as a histogram.

Specifically for NVM simulation, processing simulation data can be particularly useful for wear-leveling development and optimization. This approach allows evaluating wear-leveling techniques by simulating their impact on memory accesses without requiring modifications to complex structures, such as an operating system service. Additionally, hardware-based solutions can be assessed using a trace writer before investing in prototype development, enabling early-stage validation and refinement.

To provide a concrete example of what’s possible, we create an example trace writer that tracks NVM  per block.
Let's start by creating a fresh trace writer skeleton. Then, define a map that associates each block index with an integer value. For the sake of this example, use blocks of size 4KB. Next, within the function \texttt{SetNextAccess()}, check whether each memory access is a write operation. If it is, determine the affected block and increment its recorded access count by one.
Finally, to store the recorded data in the trace file at the end of the simulation, write the map to the trace writer’s destructor using the format \texttt{“Index | Accesses”}. Refer to \cref{lst:blockAccesses} for comparison.

\begin{lstlisting}[caption={Trace Block Accesses},label=lst:blockAccesses,language=c++,basicstyle=\ttfamily\scriptsize]
/*
 * Variables accessMap and blockSize are defined in the header file.
 * The blockSize in the example is 4KB.
 */

TutorialTraceWriter::~TutorialTraceWriter() {
    for(auto entry : accessMap) {
        traceFile << entry.first << " | " << entry.second << std::endl;
    }
}

bool TutorialTraceWriter::SetNextAccess(TraceLine *nextAccess) {
    if(nextAccess->GetOperation() == NVM::WRITE) {
        accessMap[nextAccess->GetAddress().GetPhysicalAddress() / blockSize]++;
    }

    return true
}
\end{lstlisting}

The toolchain’s repository includes two additional exercises on creating a new trace writer. For more details, please refer to Appendix~\cref{app:tudo:exercises}.

\begin{insightbox}
In this case study, we guide the reader on integrating a basic trace writer into NVMain2.0, enabling them to perform basic logging operations on the application's behavior. In addition, we show how the trace writer can be used to process simulation data for additional insight, promoting its versatility for development and analysis.
\end{insightbox}
\definecolor{hypnosVol}{HTML}{ffefba}
\definecolor{hypnosNonVol}{HTML}{87bfc7}

\subsection{Heterogeneous Cache Simulation} \label{sec:nvm-cache}

\subsubsection{Overview}

Particularly for the area of embedded systems, the SRAM technology conventionally used for cache memories has a number of sub-optimal properties.
Namely, SRAM cells are very space inefficient and incorporate a high static power consumption.
As mentioned in \cref{sec:commercial_maturity}, non-volatile STT-RAM is a promising alternative to SRAM, due to its low static power and high density allowing for smaller chip designs.
Nevertheless, a pure STT-RAM cache will likely suffer from high write overhead and cost.
Instead, a hybrid cache combining both SRAM and STT-RAM is a promising direction \cite{4798259,5090762,7479181,9399236}.  In this case study, we present a tool to evaluate different configurations of hybrid cache designs.

\subsubsection{Objective}
\begin{figure}[h]
    \begin{tikzpicture}[scale=0.4]
\path (11.75,-1.65) node[align = center] (dram) {Main\\Memory};
\path (8.25,-1.65) node[align = center] (cache) {Cache};
\path (4.75,-1.65) node[align = center] (cpu) {CPU};

\draw[fill = hypnosNonVol] (11,-0.5) rectangle (12.5,3.5);
\draw[fill = hypnosVol] (4,0.5) rectangle (5.5,2.5);

\draw[fill = hypnosVol] (7.5,1.5) rectangle (9,3);
\draw[fill = hypnosNonVol] (7.5,0) rectangle (9,1.5);

\draw[<->, line width = 1.25 pt] (5.5,1.5)--(7.5,1.5);
\draw[<->, line width = 1.25 pt] (9,1.5)--(11,1.5);

\draw[fill=hypnosVol] (-5,0.25) rectangle (-4,1.25) node[align = left, below right] () {Volatile};
\draw[fill=hypnosNonVol] (-5,1.75) rectangle (-4,2.75) node[align = left, below right] () {Non-Volatile};

\end{tikzpicture}
    \caption{Overview of the hybrid volatile/non-volatile memory hierarchies analyzed in this case study.}
    \label{fig:FAU-archOverview}
\end{figure}
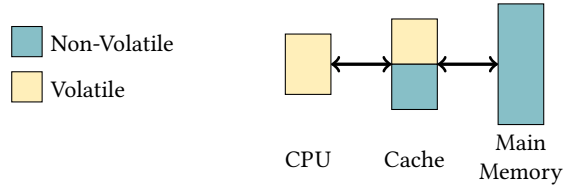

\cref{fig:FAU-archOverview} illustrates the memory hierarchy of the analyzed system, with a PCM-based main memory connected to the CPU via a hybrid cache.
This study examines how the SRAM-to-STT-RAM ratio in the hybrid cache affects dynamic energy consumption and overall latency across various embedded applications.
The goal is to provide insights for designing an optimal ratio tailored to specific application requirements.

\subsubsection{Methodology}

As gem5 does not support any hybrid caches out of the box, the following extensions were necessary.
a) STT-RAM cells have different access latencies, depending on whether a cell is read or written, with write latencies being higher as the orientation of the magnetic field has to be changed when switching the content of a cell.
Therefore, the simulator naturally had to be extended to feature asymmetric STT-RAM access latencies, along the (symmetric) access latency for SRAM cells.
b) Furthermore, the simulator was extended to generate a number of statistics specific to hybrid caches in order to give key insights into the experimental results.
Most importantly, we have added statistics to count the number of accesses, further split between read and write accesses, to each of the two cache sections, namely the volatile SRAM and the non-volatile STT-RAM section.
This access distribution can not only be used for the analysis of cache access patterns, but is also used to calculate the dynamic energy consumption caused by accesses to the different sections of the cache.
\par
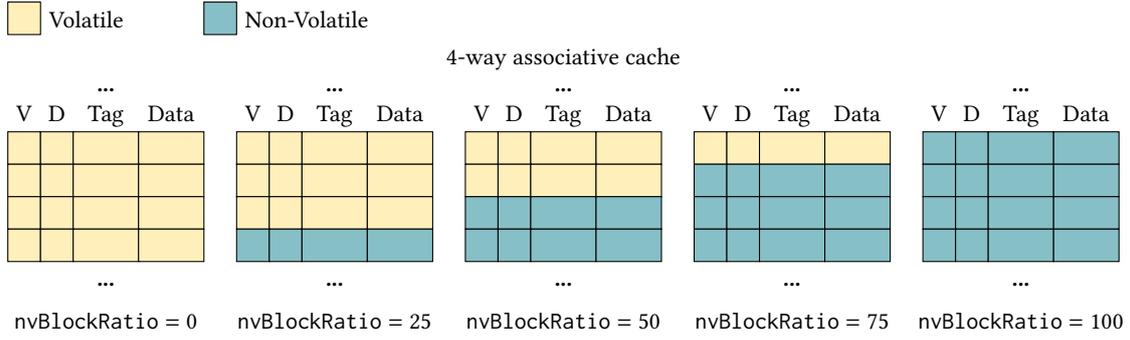
\begin{figure}
    \begin{tikzpicture}[scale = 0.43]

\path (1+14,7.25) node[align = center] () {4-way associative cache};

\path (1,6.3) node[align = center] () {\textbf{...}};
\path (1,0.3) node[align = center] () {\textbf{...}};

\path (-1.5,5.5) node[align = center] (vb) {\phantom{|}V\phantom{|}};
\path (-0.5,5.5) node[align = center] (db) {\phantom{|}D\phantom{|}};
\path (1,5.5) node[align = center] (tag) {\phantom{|}Tag\phantom{|}};
\path (3,5.5) node[align = center] (data) {\phantom{|}Data\phantom{|}};

\draw [fill = hypnosVol]  (-2,4) rectangle (-1,5);
\draw [fill = hypnosVol] (-1,4) rectangle (0,5);
\draw [fill = hypnosVol] (0,4) rectangle (2,5);
\draw [fill = hypnosVol]  (2,4) rectangle (4,5);

\draw [fill = hypnosVol]  (-2,3) rectangle (-1,4);
\draw [fill = hypnosVol] (-1,3) rectangle (0,4);
\draw [fill = hypnosVol]  (0,3) rectangle (2,4);
\draw [fill = hypnosVol]  (2,3) rectangle (4,4);

\draw [fill = hypnosVol]  (-2,2) rectangle (-1,3);
\draw [fill = hypnosVol]  (-1,2) rectangle (0,3);
\draw [fill = hypnosVol]  (0,2) rectangle (2,3);
\draw [fill = hypnosVol]  (2,2) rectangle (4,3);

\draw [fill = hypnosVol]  (-2,1) rectangle (-1,2);
\draw [fill = hypnosVol]  (-1,1) rectangle (0,2);
\draw [fill = hypnosVol]  (0,1) rectangle (2,2);
\draw [fill = hypnosVol]  (2,1) rectangle (4,2);


\path (1+7,6.3) node[align = center] () {\textbf{...}};
\path (1+7,0.3) node[align = center] () {\textbf{...}};

\path (5.5,5.5) node[align = center] (vb) {\phantom{|}V\phantom{|}};
\path (6.5,5.5) node[align = center] (db) {\phantom{|}D\phantom{|}};
\path (8,5.5) node[align = center] (tag) {\phantom{|}Tag\phantom{|}};
\path (10,5.5) node[align = center] (data) {\phantom{|}Data\phantom{|}};

\draw [fill = hypnosVol]  (5,4) rectangle (6,5);
\draw [fill = hypnosVol] (6,4) rectangle (7,5);
\draw [fill = hypnosVol] (7,4) rectangle (9,5);
\draw [fill = hypnosVol]  (9,4) rectangle (11,5);

\draw [fill = hypnosVol]  (5,3) rectangle (6,4);
\draw [fill = hypnosVol] (6,3) rectangle (7,4);
\draw [fill = hypnosVol]  (7,3) rectangle (9,4);
\draw [fill = hypnosVol]  (9,3) rectangle (11,4);

\draw [fill = hypnosVol]  (5,2) rectangle (6,3);
\draw [fill = hypnosVol]  (6,2) rectangle (7,3);
\draw [fill = hypnosVol]  (7,2) rectangle (9,3);
\draw [fill = hypnosVol]  (9,2) rectangle (11,3);

\draw [fill = hypnosNonVol]  (5,1) rectangle (6,2);
\draw [fill = hypnosNonVol]  (6,1) rectangle (7,2);
\draw [fill = hypnosNonVol]  (7,1) rectangle (9,2);
\draw [fill = hypnosNonVol]  (9,1) rectangle (11,2);


\path (1+14,6.3) node[align = center] () {\textbf{...}};
\path (1+14,0.3) node[align = center] () {\textbf{...}};

\path (12.5,5.5) node[align = center] (vb) {\phantom{|}V\phantom{|}};
\path (13.5,5.5) node[align = center] (db) {\phantom{|}D\phantom{|}};
\path (15,5.5) node[align = center] (tag) {\phantom{|}Tag\phantom{|}};
\path (17,5.5) node[align = center] (data) {\phantom{|}Data\phantom{|}};

\draw [fill = hypnosVol]  (12,4) rectangle (13,5);
\draw [fill = hypnosVol] (13,4) rectangle (14,5);
\draw [fill = hypnosVol] (14,4) rectangle (16,5);
\draw [fill = hypnosVol]  (16,4) rectangle (18,5);

\draw [fill = hypnosVol]  (12,3) rectangle (13,4);
\draw [fill = hypnosVol] (13,3) rectangle (14,4);
\draw [fill = hypnosVol]  (14,3) rectangle (16,4);
\draw [fill = hypnosVol]  (16,3) rectangle (18,4);

\draw [fill = hypnosNonVol]  (12,2) rectangle (13,3);
\draw [fill = hypnosNonVol]  (13,2) rectangle (14,3);
\draw [fill = hypnosNonVol]  (14,2) rectangle (16,3);
\draw [fill = hypnosNonVol]  (16,2) rectangle (18,3);

\draw [fill = hypnosNonVol]  (12,1) rectangle (13,2);
\draw [fill = hypnosNonVol]  (13,1) rectangle (14,2);
\draw [fill = hypnosNonVol]  (14,1) rectangle (16,2);
\draw [fill = hypnosNonVol]  (16,1) rectangle (18,2);


\path (1+21,6.3) node[align = center] () {\textbf{...}};
\path (1+21,0.3) node[align = center] () {\textbf{...}};

\path (19.5,5.5) node[align = center] (vb) {\phantom{|}V\phantom{|}};
\path (20.5,5.5) node[align = center] (db) {\phantom{|}D\phantom{|}};
\path (22,5.5) node[align = center] (tag) {\phantom{|}Tag\phantom{|}};
\path (24,5.5) node[align = center] (data) {\phantom{|}Data\phantom{|}};

\draw [fill = hypnosVol]  (7+12,4) rectangle (7+13,5);
\draw [fill = hypnosVol] (7+13,4) rectangle (7+14,5);
\draw [fill = hypnosVol] (7+14,4) rectangle (7+16,5);
\draw [fill = hypnosVol]  (7+16,4) rectangle (7+18,5);

\draw [fill = hypnosNonVol]  (7+12,3) rectangle (7+13,4);
\draw [fill = hypnosNonVol] (7+13,3) rectangle (7+14,4);
\draw [fill = hypnosNonVol]  (7+14,3) rectangle (7+16,4);
\draw [fill = hypnosNonVol]  (7+16,3) rectangle (7+18,4);

\draw [fill = hypnosNonVol]  (7+12,2) rectangle (7+13,3);
\draw [fill = hypnosNonVol]  (7+13,2) rectangle (7+14,3);
\draw [fill = hypnosNonVol]  (7+14,2) rectangle (7+16,3);
\draw [fill = hypnosNonVol]  (7+16,2) rectangle (7+18,3);

\draw [fill = hypnosNonVol]  (7+12,1) rectangle (7+13,2);
\draw [fill = hypnosNonVol]  (7+13,1) rectangle (7+14,2);
\draw [fill = hypnosNonVol]  (7+14,1) rectangle (7+16,2);
\draw [fill = hypnosNonVol]  (7+16,1) rectangle (7+18,2);

\path (1+28,6.3) node[align = center] () {\textbf{...}};
\path (1+28,0.3) node[align = center] () {\textbf{...}};

\path (26.5,5.5) node[align = center] (vb) {\phantom{|}V\phantom{|}};
\path (27.5,5.5) node[align = center] (db) {\phantom{|}D\phantom{|}};
\path (29,5.5) node[align = center] (tag) {\phantom{|}Tag\phantom{|}};
\path (31,5.5) node[align = center] (data) {\phantom{|}Data\phantom{|}};

\draw [fill = hypnosNonVol]  (7+7+12,4) rectangle (7+7+13,5);
\draw [fill = hypnosNonVol] (7+7+13,4) rectangle (7+7+14,5);
\draw [fill = hypnosNonVol] (7+7+14,4) rectangle (7+7+16,5);
\draw [fill = hypnosNonVol]  (7+7+16,4) rectangle (7+7+18,5);

\draw [fill = hypnosNonVol]  (7+7+12,3) rectangle (7+7+13,4);
\draw [fill = hypnosNonVol] (7+7+13,3) rectangle (7+7+14,4);
\draw [fill = hypnosNonVol]  (7+7+14,3) rectangle (7+7+16,4);
\draw [fill = hypnosNonVol]  (7+7+16,3) rectangle (7+7+18,4);

\draw [fill = hypnosNonVol]  (7+7+12,2) rectangle (7+7+13,3);
\draw [fill = hypnosNonVol]  (7+7+13,2) rectangle (7+7+14,3);
\draw [fill = hypnosNonVol]  (7+7+14,2) rectangle (7+7+16,3);
\draw [fill = hypnosNonVol]  (7+7+16,2) rectangle (7+7+18,3);

\draw [fill = hypnosNonVol]  (7+7+12,1) rectangle (7+7+13,2);
\draw [fill = hypnosNonVol]  (7+7+13,1) rectangle (7+7+14,2);
\draw [fill = hypnosNonVol]  (7+7+14,1) rectangle (7+7+16,2);
\draw [fill = hypnosNonVol]  (7+7+16,1) rectangle (7+7+18,2);


\draw[fill=hypnosVol] (-2,8) rectangle (-1,9) node[align = left, below right] () {Volatile};
\draw[fill=hypnosNonVol] (4,8) rectangle (5,9) node[align = left, below right] () {Non-Volatile};

\path (1,-.9) node[align = center] (data) {\phantom{|}$\text{\texttt{nvBlockRatio}}=0$\phantom{|}};

\path (8,-.9) node[align = center] (data) {\phantom{|}$\text{\texttt{nvBlockRatio}}=25$\phantom{|}};

\path (8+7,-.9) node[align = center] (data) {\phantom{|}$\text{\texttt{nvBlockRatio}}=50$\phantom{|}};

\path (8+14,-.9) node[align = center] (data) {\phantom{|}$\text{\texttt{nvBlockRatio}}=75$\phantom{|}};

\path (8+21,-.9) node[align = center] (data) {\phantom{|}$\text{\texttt{nvBlockRatio}}=100$\phantom{|}};

\end{tikzpicture}
    \caption{Hybrid (mixed volatile/non-volatile) cache architecture. Visualization on how the \texttt{nvBlockRatio} Parameter influences the technological composition of a cache set.}
    \label{fig:FAU-nvBlockratio}
\end{figure}

As mentioned in the previous section, the degree to which non-volatility is introduced to the cache hierarchy is a key point of our analysis.
To be precise, the hybridization of the caches takes place on cache set level.
The cache set where requested data is potentially located is determined by the address of the request.
In an $n$-way associative cache, implementing $n_0$ cache lines per set in SRAM and $n_1$ cache lines in STT-RAM technology, wrt. $n_0 + n_1 = n$, thus allows for data to be placed in either the SRAM or the STT-RAM section.
In order to evaluate hybrid caches with different degrees of non-volatility, we have added a parameter called \texttt{nvBlockRatio} which sets the percentage of cache lines per set that are simulated as being implemented in the non-volatile STT-RAM technology, i.e., $n_1 = \left\lfloor\frac{\text{\texttt{nvBlockRatio}}}{100} \cdot n\right\rfloor$.
The effect of different \texttt{nvBlockRatio} settings on the configuration can be seen in \cref{fig:FAU-nvBlockratio}
\par
Using the Unikraft, we build the unikernels containing our test applications for this case study:
a) An \emph{image processing application} performing 2D convolution on a $384 \times 384$ large input image using a $3 \times 3$ large kernel and
b) a \emph{merge sort application} sorting an array consisting of $32,768$ integers.
These applications are chosen as they incorporate typical tasks for embedded systems and feature vastly different memory access patterns.
More precisely, the image processing application is on the read-intensive side, with 9 input image values and 9 kernel values required to be read in order to write a single output image value.
Contrarily, the merge sort application can be located on the write-intensive side.
This is due to the sub-arrays generated after each split of the first phase of the merge sort algorithm being written to a new memory location, before writing back the sorted values to the original array in the second phase of the algorithm.
The compiled unikernels can then simply be handed to the simulator.
\par
The latency and energy parameters selected for the SRAM and STT-RAM cache sections, respectively, are of utmost importance for the validity of any experimental results.
As accurate parameters stemming from measurements on actual hardware are hard to come by, along with STT-RAM based caches still not yet being commercially available on a widespread basis, we rely on NVSim \cite{6218223} to provide us with the necessary parameters on different memory technologies.
The selected parameters are displayed in \cref{tab:FAU-param}.
\begin{table}
\renewcommand{\cellalign}{cr}
\begin{tabular}{lrrrr}
\toprule
		& \textbf{Read Latency} &  \textbf{Write Latency} & \makecell{\textbf{Read Energy}\\\textbf{(per access)}} & \makecell{\textbf{Write Energy}\\\textbf{(per access)}}\\ \midrule
\textbf{SRAM Cache} & 2 Cycles @240 MHz & 2 Cycles @240 MHz & 0.009 nJ & 0.009 nJ\\
\textbf{STT-RAM Cache} & 2 Cycles @240 MHz & 8 Cycles @240 MHz & 0.007 nJ & 0.056 nJ\\
\bottomrule
\end{tabular}
\caption{Characterized read/write latencies and average read/write access energies of the SRAM and STT-RAM cache section, respectively.}
\label{tab:FAU-param}
\end{table}
However, since the toolchain is highly flexible, plugging in different parameter values, in case more accurate measurements become available, is straightforward and does not require any changes to the underlying framework.
NVMain2.0 is further utilized to simulate the underlying non-volatile PCM main memory modeled after the characteristics given in \cite{6176872}.
\par
\begin{figure}
	\begin{subfigure}{0.475\textwidth}
		\begin{tikzpicture}
 
\begin{axis}[
	xmin = -10,
	xmax = 110,
	xtick distance = 25,
	ymin = 0,
	ymax = 20000000,
	ytick distance = 5000000,
    grid = both,
    major grid style = {gray!20},
    width = \columnwidth,
    height = 0.78\textwidth,
    xlabel = { \normalsize Degree of non-volatility (in \%)},
    ylabel = {\small Latency (in CPU cycles)},
    scale = 0.72
    ]

\addplot[red,mark=*,mark options={scale=1.25, fill=blue},text mark as node=true,point meta=explicit symbolic] coordinates { (0,15904185) (25,15904185) (50,15904185) (75,15904185) (100,15904185)};

\end{axis}
\end{tikzpicture}
	\end{subfigure}
	\begin{subfigure}{0.475\textwidth}
		\begin{tikzpicture}
 
\begin{axis}[
	xmin = -10,
	xmax = 110,
	xtick distance = 25,
	ymin = 0,
	ymax = 100000,
	ytick distance = 25000,
    grid = both,
    major grid style = {gray!20},
    width = \columnwidth,
    height = 0.78\textwidth,
    xlabel = { \normalsize Degree of non-volatility (in \%)},
    ylabel = {\small Dynamic Cache Energy (in nJ)},
    scale = 0.72
    ]

\addplot[red,mark=*,mark options={scale=1.25, fill=blue},text mark as node=true,point meta=explicit symbolic] coordinates { (0,23217.578001) (25,37583.014876) (50,52414.414793) (75,66654.849654) (100,80393.959481)};

\end{axis}
\end{tikzpicture}
	\end{subfigure}
    \caption{Latency and dynamic energy consumption under different degrees of non-volatility for a write-intensive merge sort application.}
    \label{fig:FAUmsort}
\end{figure}
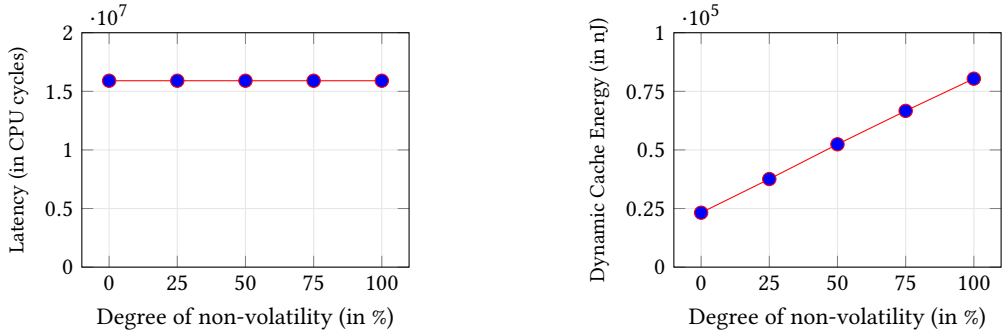
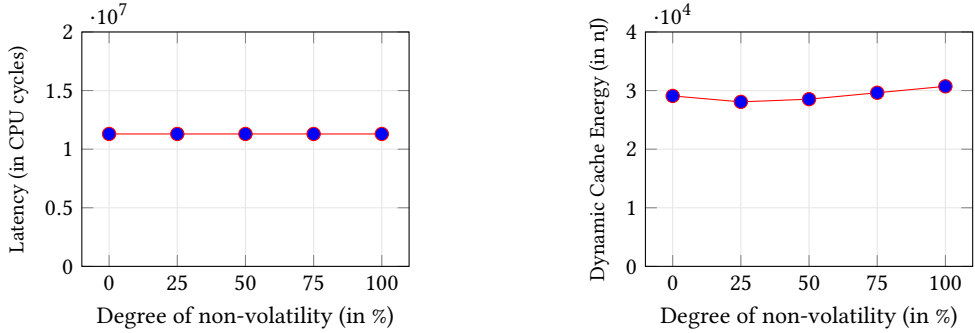
\begin{figure}
	\begin{subfigure}{0.475\textwidth}
		\begin{tikzpicture}
 
\begin{axis}[
	xmin = -10,
	xmax = 110,
	xtick distance = 25,
	ymin = 0,
	ymax = 20000000,
	ytick distance = 5000000,
    grid = both,
    major grid style = {gray!20},
    width = \columnwidth,
    height = 0.78\textwidth,
    xlabel = { \normalsize Degree of non-volatility (in \%)},
    ylabel = {\small Latency (in CPU cycles)},
    scale = 0.72
    ]

\addplot[red,mark=*,mark options={scale=1.25, fill=blue},text mark as node=true,point meta=explicit symbolic] coordinates { (0,11298120) (25,11298143) (50,11298143) (75,11298118) (100,11298118)};

\end{axis}
\end{tikzpicture}
	\end{subfigure}
	\begin{subfigure}{0.475\textwidth}
		\begin{tikzpicture}
 
\begin{axis}[
	xmin = -10,
	xmax = 110,
	xtick distance = 25,
	ymin = 0,
	ymax = 40000,
	ytick distance = 10000,
    grid = both,
    major grid style = {gray!20},
    width = \columnwidth,
    height = 0.78\textwidth,
    xlabel = { \normalsize Degree of non-volatility (in \%)},
    ylabel = {\small Dynamic Cache Energy (in nJ)},
    scale = 0.72
    ]

\addplot[red,mark=*,mark options={scale=1.25, fill=blue},text mark as node=true,point meta=explicit symbolic] coordinates { (0,29088.925748) (25,28074.614715) (50,28530.362257) (75,29628.430611) (100,30723.784948)};

\end{axis}
\end{tikzpicture}
	\end{subfigure}
    \caption{Latency and dynamic energy consumption under different degrees of non-volatility for a read-intensive image processing application.}
    \label{fig:FAUip}
\end{figure}

In the following, we run the simulation for a system with a 240 MHz out-of-order CPU, employing a 32 kB 4-way-associative hybrid data cache.
By increasing the \texttt{nvBlockRatio} Parameter from 0 to 100 in steps of 25 (acc.\ to \cref{fig:FAU-nvBlockratio}), 
we can observe the results displayed in \cref{fig:FAUmsort} and \cref{fig:FAUip}.

\textit{Latency:} We can identify, that a higher degree of non-volatility barely influences the overall latency, despite the high NVM write latency.
This is explained by the fact, that in case of a cache write the CPU does not stall until the write is completed, unless data from the very same cache line is required in the meantime, which is rarely the case.

\textit{Dynamic Energy:} While the observation in the latency dimension holds for both applications, the dynamic energy under varying degrees of non-volatility behaves differently for the two applications.
For the write-intensive merge sort application, displayed in \cref{fig:FAUmsort}, the dynamic energy consumption rises proportionally with the increased degree of non-volatility.
This is to be expected, as the higher degree of non-volatility also implies a higher number of high energy NVM writes.
However, this is not the case for the image processing experiment, as seen in \cref{fig:FAUip}.
Here, as the application is highly read-intensive with read accesses in STT-RAM being more energy efficient than in SRAM, introducing NVM to a low degree can even reduce the dynamic energy consumption of the cache.
Using the toolchain with our hybrid cache extensions, we can thus analyze in which application scenarios and to which degree the introduction of NVM to the cache hierarchy proves the most beneficial.
Furthermore, for this case study, no replacement policies that consider the characteristics of the underlying memory technologies were analyzed.
These policies performing placement decisions according to, e.g., the predicted write intensity of future accesses, can further boost the performance of hybrid caches to a notable degree \cite{8716297,7110550, Wilbert:2024a}.

\subsubsection{Step-by-Step Instructions}
In the following we will provide a step-by-step guide on how hybrid caches are implemented in gem5.
First, we will create a \texttt{HybridCache} \texttt{SimObject} by inheriting the base methods from gem5's base cache class located in \texttt{src/mem/cache/base.cc}.
To account for the hybrid cache's asymmetric read/write latencies, we add additional parameters for the read and write latency of the non-volatile STT-RAM section.
Additionally, in order to generate meaningful statistics that allow for the analysis of hybrid caches, we add statistics for the dynamic energy consumption and number of accesses per section.
The gem5 \texttt{CacheBlk}, representing a cache line, is extended to a \texttt{HybridCacheBlk}, carrying the additional information whether the cache line is considered to be implemented in a volatile or non-volatile manner.
Whenever a cache line is accessed in the extended base cache class, we check for the volatility of the corresponding \texttt{HybridCacheBlk}.
Depending on whether the cache line is volatile or not, we thus take the latency and energy parameters for the SRAM or the STT-RAM section, respectively, as seen in the following listing where we exemplarily display the code for setting the latency and logging statistics following a write access to the hybrid cache.
\begin{lstlisting}[language=C]
HybridCacheBlk* hyblk = dynamic_cast<HybridCacheBlk*>(blk);
Cycles lat = hyblk->isVolatile() ? dataLatency : dataWriteLatency;
if (hyblk->isVolatile()) {
	hybrid_stats.noOfVolWrites++;
	hybrid_stats.dynEnergy += volWriteEnergy;
} else {
	hybrid_stats.noOfNonVolWrites++;
	hybrid_stats.dynEnergy += nonVolWriteEnergy;
}
\end{lstlisting}
The cache lines themselves are initialized and managed in the \texttt{BaseTags} class located in \texttt{src/mem/cache/tags/base.cc}.
To set the cache lines as volatile/non-volatile in accordance to the previously described \texttt{nvBlockRatio} parameter, we set the \texttt{HybridCacheBlk}'s volatile flag in the \texttt{HybridSetAssoc} class, which is derived from the \texttt{BaseTags} class.
\begin{lstlisting}[language=C]
int numNvBlocksPerSet = (nvBlockRatio/100)*assoc;
int numNvBlocksInSet = 0;
// Initialize all blocks
for (int blk_index = 0; blk_index < numBlocks; blk_index++) {
    HybridCacheBlk* blk = &blks[blk_index];
    if (blk_index 
        numNvBlocksInSet = 0;
    // Set whether block is volatile or non-volatile
    if (numNvBlocksInSet < numNvBlocksPerSet) {
        blk->setVolatile(false);
        numNvBlocksInSet++;
    } else {
        blk->setVolatile(true);
    }
}
\end{lstlisting}
On the Python configuration level in \texttt{configs/common/Caches.py}, we can define, e.g., L1 Data Caches as shown below, which are derived from the introduced \texttt{HybridCache} class, with hybrid-cache-specific parameters set to any desired value.
\begin{lstlisting}[language = Python]
class L1DCache(HybridCache):
    assoc = 4
    nv_block_ratio = 50
    data_read_latency = 2
    data_write_latency = 8
    ...
\end{lstlisting}
The connection between the C++ \texttt{SimObjects} containing the functional implementation and the Python classes setting a configuration is realized in \texttt{src/mem/cache/Cache.py} as follows.
\begin{lstlisting}[language = Python]
class HybridCache(ClockedObject):
    type = 'HybridCache'
    cxx_header = 'mem/cache/hybrid_cache.hh'
    cxx_class = 'gem5::HybridCache'
    data_read_latency = Param.Cycles("Data read access latency")
    data_write_latency = Param.Cycles("Data write access latency")
    ...
\end{lstlisting}
Furthermore, the additional hybrid-cache-specific parameters which can be set in the configuration files are named here.
For the \texttt{HybridSetAssoc} class, this is performed analogously in the \texttt{src/mem/cache/tags/Tags.py} file.
By running a gem5 simulation with Cache \texttt{SimObjects} of the \texttt{HybridCache} type, the analysis of heterogeneous cache architectures is thus made possible.
See \cref{app:fau}, for a more detailed description of the concrete command to perform and evaluate a simulation run.
\begin{insightbox}
In this case study, we have learned how gem5 can be extended to support caches incorporating both SRAM and STT-RAM cells.
We have observed in the experiments, that the write latency overhead of STT-RAM is barely noticeable in application scenarios.
While the introduction of NVM inherently lowers static power consumption in caches, even under conventional replacement policies, hybrid caches can further reduce dynamic energy consumption, depending on the characteristics of the application.
Owing to gem5's flexibility, our extension thus provides the opportunity to quickly evaluate the potential of hybrid caches when developing novel cache policies or investigating novel use-cases.

\end{insightbox}

\subsection{Compute-in-Memory (CiM)} \label{sec:nvm-cim}

\subsubsection{Overview}

Modern computing systems predominantly rely on the von Neumann architecture \cite{kit1}, where the CPU handles data processing and memory components are dedicated to storage \cite{kit2}. This design results in frequent data transfers between the CPU and memory, causing significant energy consumption and performance bottlenecks, collectively known as the ``memory wall.'' These transfers account for up to 60\% of system energy use, with memory access consuming far more energy than computational operations \cite{kit3,kit4,kit5}. Additionally, DRAM faces scaling limitations, with its capacity growth lagging behind increases in processing power, partly due to the shrinking reliability of DRAM cells \cite{kit/dramScale1,kit/dramScale2,kit/dramScale3}.

The Compute-in-Memory (CiM) architecture has emerged as a viable solution to reduce the energy cost of data movement. By bringing processing capabilities closer to the data, CiM reduces reliance on the CPU \cite{kit/ext1,kit/ext2,kit/ext3}. These architectures range from minimal hardware changes for simple operations, such as bulk data manipulation, to more complex designs integrating processing cores near memory arrays. While the latter supports a wide range of applications, challenges such as design complexity, frequent CPU interactions, and task-offloading inefficiencies limit their practicality \cite{kit/ext2}. Simplified CiM designs, focusing on specific applications, reduce information exchange with the CPU and ease programming challenges, making them ideal for parallelizable tasks like DNA sequencing \cite{kit/dna1,kit/dna2, hameed_tetc21}, image processing \cite{kit/img1,kit/img2}, and database operations \cite{kit/db1,kit/db2}.
As advanced memory technologies that tightly couple memory with logic units have entered mass production (e.g., HBM \cite{kit/hbm} and HMC \cite{kit/hmc}), CiM is the industry's next frontier~\cite{khan_cimlandscape_2024}.

However, current research has mostly focused on SRAM-based and DRAM-based CiM architectures, leaving a gap in tools for NVM-based CiM design and evaluation. NVMs, as discussed in \cref{sec:Introduction}, offer significant advantages such as reduced static power leakage, no need for data refreshing, and improved scalability. These properties make NVMs particularly promising for CiM, where energy efficiency and scalability are critical \cite{kit/ext1,kit/ext4}. NVM like ReRAM has resistive properties that enable efficient analog-based MAC and Boolean computations with minimal hardware modifications.

Recent advancements in NVM-based CiM modules demonstrate their potential for practical implementations. Leading foundries, including TSMC, Samsung, and IBM, have fabricated CiM modules using ReRAM, STT-MRAM, and PCM. For instance, TSMC has integrated ReRAM and STT-MRAM into CiM designs for neural network acceleration \cite{kit/stt-tsmc1,kit/reram-tsmc4}, while Samsung and IBM have demonstrated CiM modules using STT-MRAM and PCM \cite{kit/pcm-ibm3,kit/stt-samsung2}, respectively. These developments showcase the ability of NVM-based CiM architectures to address the performance and energy limitations of traditional computing systems, highlighting their potential to enable more efficient and scalable designs.

\subsubsection{Objective}

Existing CiM simulators, such as CIM-SIM \cite{CIM-SIM}, MNSIM \cite{MNSIM}, NeuroSiM \cite{NEUROSIM}, PiMulator \cite{PiMulator}, MultiPIM \cite{MultiPIM}, Sim2PIM \cite{Sim2PIM}, and PIMSim \cite{PIMSim}, offer capabilities like full-system simulation, reconfigurable technology and architecture, and cycle-accurate modeling.
However, our CiM extension is the first to integrate all these features specifically for NVM.
Designed with detailed modeling, modularity, and extensibility,
our extension prioritizes realistic design choices to ensure practical applicability in real-world scenarios.
The following subsections provide an in-depth exploration of these features.

\subsubsection{Methodology}

To gain a deeper understanding of the proposed CiM extension, we begin by discussing the general design choices, the available alternatives, and the reasoning behind these decisions. This approach helps readers comprehend the requirements for practical architectural design while enabling them to identify potential improvements and trade-offs. Subsequently, we provide a detailed implementation guide for the gem5 framework in the following subsections.

\textbf{CiM vs. PIM:}
A central distinction in this design is between CiM and PIM (Processing-in-Memory). PIM integrates processing cores near memory to support diverse applications but faces challenges such as complex design, frequent CPU data exchanges, and system support issues (e.g., coherency mechanisms and task-offloading granularity)  \cite{kit/ext1,kit/ext4,kit/ext5,kit/ext7}.
CiM, by contrast, focuses on minimal modifications to the memory periphery for specific tasks like bulk bitwise operations or data copying \cite{kit/rowclone}. This narrower scope simplifies programming, reduces CPU interaction, and enables a streamlined design. Additionally, CiM leverages the unique resistive properties of NVM for efficient analog computations, such as MAC and Boolean operations \cite{kit/ext4,kit/ext5,kit/ext8}, using simple crossbar circuitry modifications. This design resembles SIMD cores with much larger register sizes (equivalent to a memory bank’s row), making CiM highly efficient for parallelized, data-intensive applications.

\textbf{CPU-CiM Communication:}
This can be implemented in two ways:
\CRicon{1} Custom Instructions and \CRicon{2} Memory-Mapped I/O. While the The first approach is well-suited for accelerators closely connected to the CPU core. However, it requires significant CPU modifications (e.g., pipeline and control logic changes) and is tied to specific ISAs, which limits compatibility.
The second approach maps the CiM module’s physical address space into the virtual address space of applications. While it introduces communication delays due to OS involvement, it avoids the design challenges of custom instructions and is more flexible, making it the preferred choice in this case.

\textbf{Physical Placement of CiM Components:}
A critical design decision for CiM components is their physical placement within the main memory module, which consists of multiple memory chips. While integrating CiM circuitry directly into each chip allows for parallel processing of simple operations like logical AND, it faces challenges with operations like logical shifts that require contiguous data. Modeling CiM as a separate chip adjacent to memory chips addresses some issues but introduces power, latency, and communication protocol challenges due to the DDR interface’s lack of dedicated signal lines. Despite these trade-offs, using LRDIMM (Load-Reduced Dual In-Line Memory Module) modules, commonly found in server applications, offers a practical solution. LRDIMMs enhance signal integrity, reduce system modifications, and provide added benefits like error correction and improved simulation flexibility, making them ideal for CiM integration.

\begin{figure}[ht]
	\centering
	\begin{minipage}[c]{0.4\textwidth}
		\centering
		\includegraphics[width=\textwidth, page=1]{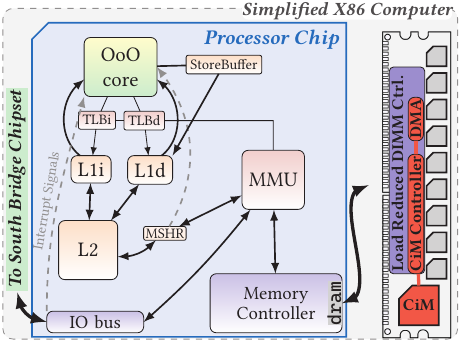} 
		\caption{x86 with CiM-capable memory}
		\label{kit:fig1a}
	\end{minipage}
	\hfill
	\begin{minipage}[c]{0.49\textwidth}
		\centering
		\includegraphics[width=\textwidth, page=2]{KIT-figures/mainSys.pdf} 
		\caption{Structure of the CiM chip}
		\label{kit:fig1b}
	\end{minipage}
\end{figure}

Finally, memory controllers, which are highly complex and integrated into modern CPUs, present a unique challenge for CiM implementation. Modifying these controllers requires careful consideration to avoid performance bottlenecks or compatibility issues. Proposals involving such changes must carefully balance hardware and software implications.
As with other extensions discussed in this article, we utilize the gem5 framework \cite{kit/gem5-1,kit/gem5-2} for modeling and simulation purposes.
\cref{kit:fig1a} shows a simplified X86 system where minimal changes are required for CiM integration. Only a few components within the LRDIMM module are modified, while the CPU-integrated memory controller and memory chips remain unchanged.

\subsubsection{Step-by-Step Instructions}\label{kit/label1}

\begin{figure}[h]
	\centering
	\includegraphics[width=0.195\linewidth, page=1]{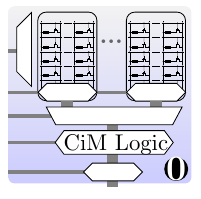}
	\includegraphics[width=0.195\linewidth, page=2]{KIT-figures/steps.pdf}
	\includegraphics[width=0.195\linewidth, page=3]{KIT-figures/steps.pdf}
	\includegraphics[width=0.195\linewidth, page=4]{KIT-figures/steps.pdf}
	\includegraphics[width=0.195\linewidth, page=5]{KIT-figures/steps.pdf}
	\caption{Steps required to perform a NAND operation.}
	\label{kit/fig2}
\end{figure}

This subsection discusses the programming model for the CiM extension, provides examples and supported operations, and concludes with a brief discussion on how it is enabled in gem5. The CiM extension operates similarly to a SIMD core, with CiM rows acting as registers and supporting only simple operations on column-aligned data. A CiM-capable chip resembles a conventional memory chip with modifications to row decoders (`\texttt{Dec.}', for multi-row activation) and sense amplifiers (`\texttt{SA}', for multi-reference voltages), as illustrated in \cref{kit:fig1b}. Simple operations, like bitwise AND, are performed by simultaneously activating rows and applying the appropriate reference voltage across the sense amplifiers.
To support more complex operations, CMOS-based CiM Logic is added after the column selector multiplexer, enabling tasks like bitwise NOT. Input signals for these units are managed by a state machine in the CiM controller, which processes commands from user applications. As illustrated in \cref{kit/fig2}, a NAND operation involves activating operand rows for a parallel AND (step \CRicon{1}), loading results into the CiM Logic for a NOT operation (steps \CRicon{2} to \CRicon{3}), and storing the final results back into memory (step \CRicon{4}). \cref{kit/lst1} shows the implementation of the NAND operation in a user application with the help of CiM API.
To provide a practical example that includes both a loop and a conditional statement, \cref{kit/lst2} presents a sequential \texttt{for} loop in regular C++ code that uses the ternary operator (`\texttt{?:}'), and \cref{kit/lst3} shows the conversion of this code into the CiM version, with the loop unrolled.

\cref{kit-tb1} lists the essential operations implemented in the CiM extension. We have applied these operations to several real-world applications, including the following:
\CRicon{1} BitIndexing (BIT)\cite{kit/app1}: Bitmap-formatted database queries are ORed together, and the results are ANDed with another query.
\CRicon{2} BLASTN algorithm (BLS)\cite{kit/app2}: Used for searching short DNA sequences within large DNA or protein sequences.
\CRicon{3} Morphological Image Processing (MIP)\cite{kit/app3}: Implementation of Dilation and Erosion algorithms with binary-coded input images of various filter sizes.
\CRicon{4} Marching Squares (MSQ)\cite{kit/app4}: An algorithm designed to extract contour lines from 2D images.
\CRicon{5} Shifted Hamming Distance (SHD)\cite{kit/app5}: A well-known algorithm for calculating edit distances in short DNA sequences.
\CRicon{6} BitWeaving (BWV)\cite{kit/app6}: A technique developed for efficiently scanning database queries in memory.
These applications are categorized as \textit{Embarrassingly Parallel} \cite{kit/emb}, making them well-suited for CiM implementation.
Other operations, such as MAC for binarized vector-to-matrix multiplication, are also included in the source files to support the implementation of simple neural network applications.
Users can refer to the provided application examples, CiM API code and documentation, as well as other modules, such as OS drivers offered by the CiM API, to gain a better understanding of CiM implementation.

\begin{table}[h]
	\caption{Essential operations supported by the CiM extension.}
	\label{kit-tb1}
	\begin{tabular}{@{}ccccc@{}}
		\toprule
		Operation Type                                                                                     &
		Operation                                                                                          &
		Source                                                                                             &
		Destination                                                                                        &
		Description                                                                                                                               \\ \midrule
		In Memory Array                                                                                    &
		\texttt{AND, OR, XOR}                                                                              &
		A List of Rows                                                                                     &
		\begin{tabular}[c]{@{}c@{}}A Single Row\\ (Default: \texttt{SA} output)\end{tabular}                        &
		Perform bitwise operation                                                                                                             \\ \midrule
		\multirow{2}{*}[-6pt]{In CMOS Periphery}                                                           &
		\texttt{COPY}                                                                                      &
		\multirow{2}{*}[-6pt]{\begin{tabular}[c]{@{}c@{}}A Single Row\\ (Default: \texttt{SA} output)\end{tabular}} &
		\multirow{2}{*}[-6pt]{A Single Row}                                                                &
		\begin{tabular}[c]{@{}c@{}}Copies row content\\ Optional: bitwise rotation\end{tabular}                                               \\ \cmidrule(lr){2-2} \cmidrule(l){5-5}
		                                                                                                   &
		\texttt{NOT\_COND}                                                                                 &
		                                                                                                   &
		                                                                                                   &
		\begin{tabular}[c]{@{}c@{}}Performs bitwise NOT\\ Or vector mask operation.\end{tabular}                                              \\ \midrule
		\multirow{2}{*}[-3pt]{\begin{tabular}[c]{@{}c@{}}In CiM Controller\\ Or DMA-Related\end{tabular}}        &
		\texttt{copy\_to\_cim}                                                                             &
		A pointer to a vector                                                                              &
		CiM row number                                                                                     &
		\multirow{2}{*}[-3pt]{\begin{tabular}[c]{@{}c@{}}Copies vector to/from a CiM row\\ If DMA:  Main Memory $\leftrightarrow$ CiM\end{tabular}} \\ \cmidrule(lr){3-4}
		                                                                                                   &
		\texttt{copy\_to\_cpu}                                                                             &
		CiM row number                                                                                     &
		A pointer to a vector                                                                              &
		\\ \bottomrule
	\end{tabular}
\end{table}

\begin{figure}[h]
	\centering
	\includegraphics[width=0.8\linewidth, page=3]{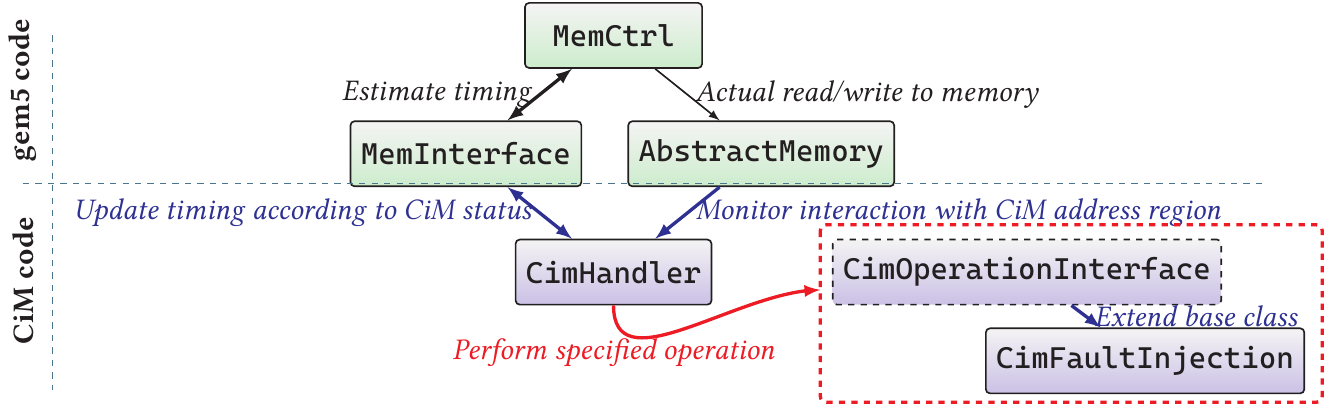}
	\caption{Summary of modifications, contributions, and communications regarding the CiM extension in gem5.}
	\label{kit/fig3}
\end{figure}

To summarize, \cref{kit/fig3} illustrates the modifications made to the gem5 framework to enable the CiM extension. The only alterations to the gem5 codebase involve the \texttt{AbstractMemory} class, which has been adapted to monitor interactions with the CiM address space, and the \texttt{MemInterface} class, which has been updated to adjust the access time for any request associated with the CiM address space. The \texttt{CiMHandler} class serves as the primary component responsible for managing CiM commands. It accomplishes this by interpreting opcodes, configuring the state machine, assigning appropriate timing, and ultimately invoking the relevant operation function via the \texttt{CiMOperationInterface}.
The \texttt{CiMOperationInterface} acts as the base class for all CiM operations, providing a pure functional implementation. This class can be extended to incorporate detailed modeling or fault injection capabilities (e.g., the \texttt{CiMFaultInjection} class, as shown). Most parameters can be configured through the gem5 simulation configuration file, allowing users to select among different technologies or specify various timing values. Additionally, users can extend the source code by overriding the provided functions.
Finally, as with the previous extensions discussed in this article, either the `\texttt{CDNCcim=1}' or `\texttt{CDNCcimFS=1}' flag must be specified during the gem5 compilation process, depending on the desired simulation mode (system emulation/full-system simulation mode), to enable the CiM extension.
Detailed steps for setting up and running the CiM extension are provided in the Appendices section.

\begin{insightbox}
	This case study provides a comprehensive guide to the practical realization of CiM, highlighting key challenges and offering researchers clear directions for getting started.
	Our CiM extension stands out by combining full-system simulation, flexible technology and architecture, and detailed circuit modeling, specifically targeting NVM technologies.
	It focuses on accuracy, reconfigurability, and expandability, enabling realistic design decisions and facilitating hardware/software design-space exploration for CiM applications.
\end{insightbox}

\section{Conclusion}
In this collaborative work from the SPP 2377 priority program, we present a tutorial that demonstrates extensions to a toolchain based on gem5 and NVMain. We show how to use the toolchain for architectural exploration and development of co-design strategies for emerging memory technologies. Four case studies were discussed: hybrid main memory, trace-based wear-out analysis, heterogeneous cache design, and compute-in-memory architectures. These case studies not only demonstrate the toolchain's versatility and depth, but also provide researchers with the knowledge they need to extend and customize it for their specific applications. We also provide open access to the source code and setup instructions in the hope of increasing the reproducibility of research in NVM co-design.

\bibliographystyle{plain}
\bibliography{references}

\begin{thebibliography}{100}

\bibitem{digikey}
Digikey.
\newblock \url{https://digikey.de}.
\newblock [Accessed 28-11-2024].

\bibitem{gem5:documentation}
gem5: {B}uilding gem5 --- gem5.org.
\newblock \url{https://www.gem5.org/documentation/general_docs/building}.
\newblock [Accessed 27-01-2025].

\bibitem{repo}
{G}it{H}ub - {O}{M}{A}-{N}{V}{M}/{N}{V}{M}\_{S}imulation: {A} simulator ({G}em5
  + {N}{V}{M}ain2.0) for {N}{V}{M} --- github.com.
\newblock \url{https://github.com/OMA-NVM/NVM_Simulation}.
\newblock [Accessed 13-01-2025].

\bibitem{docker}
{H}ome --- docker.com.
\newblock \url{https://www.docker.com/}.
\newblock [Accessed 13-01-2025].

\bibitem{vscode}
{V}isual {S}tudio {C}ode - {C}ode {E}diting. {R}edefined ---
  code.visualstudio.com.
\newblock \url{https://code.visualstudio.com/}.
\newblock [Accessed 13-01-2025].

\bibitem{vscodeextensions}
{V}isual {S}tudio {C}ode {R}emote {D}evelopment --- code.visualstudio.com.
\newblock \url{https://code.visualstudio.com/docs/remote/remote-overview}.
\newblock [Accessed 13-01-2025].

\bibitem{7110550}
Junwhan Ahn, Sungjoo Yoo, and Kiyoung Choi.
\newblock Prediction hybrid cache: An energy-efficient stt-ram cache
  architecture.
\newblock {\em IEEE Transactions on Computers}, 65(3):940--951, 2016.

\bibitem{kit/dna2}
Stephen~F Altschul, Warren Gish, Webb Miller, Eugene~W Myers, and David~J
  Lipman.
\newblock Basic local alignment search tool.
\newblock {\em Journal of molecular biology}, 215(3):403--410, 1990.

\bibitem{kit/app2}
Stephen~F Altschul, Warren Gish, Webb Miller, Eugene~W Myers, and David~J
  Lipman.
\newblock Basic local alignment search tool.
\newblock {\em Journal of molecular biology}, 215(3):403--410, 1990.

\bibitem{8351201}
Artur Antonyan, Suksoo Pyo, Hyuntaek Jung, and Taejoong Song.
\newblock Embedded mram macro for eflash replacement.
\newblock In {\em 2018 IEEE International Symposium on Circuits and Systems
  (ISCAS)}, pages 1--4, 2018.

\bibitem{ArmN1}
{Arm Ltd.}
\newblock {\em Arm Neoverse N1 CPU: Accelerating the Transformation to a
  Scalable Cloud-to-Edge Infrastructure}.
\newblock Arm Ltd., Cambridge, UK, February 2019.
\newblock Available online: \url{https://www.arm.com} (Accessed: 12 Feb 2025).

\bibitem{kit/ext5}
Kazi Asifuzzaman, Narasinga~Rao Miniskar, Aaron~R Young, Frank Liu, and
  Jeffrey~S Vetter.
\newblock A survey on processing-in-memory techniques: Advances and challenges.
\newblock {\em Memories-Materials, Devices, Circuits and Systems}, 4:100022,
  2023.

\bibitem{CIM-SIM}
Ali BanaGozar, Kanishkan Vadivel, Sander Stuijk, Henk Corporaal, Stephan Wong,
  Muath~Abu Lebdeh, Jintao Yu, and Said Hamdioui.
\newblock Cim-sim: Computation in memory simuiator.
\newblock In {\em Proceedings of the 22nd International Workshop on Software
  and Compilers for Embedded Systems}, SCOPES '19, page 1–4, New York, NY,
  USA, 2019. Association for Computing Machinery.

\bibitem{binkert:2011}
Nathan Binkert, Bradford Beckmann, Gabriel Black, Steven~K. Reinhardt, Ali
  Saidi, Arkaprava Basu, Joel Hestness, Derek~R. Hower, Tushar Krishna, Somayeh
  Sardashti, Rathijit Sen, Korey Sewell, Muhammad Shoaib, Nilay Vaish, Mark~D.
  Hill, and David~A. Wood.
\newblock The {Gem5} simulator.
\newblock {\em SIGARCH Comput. Archit. News}, 39(2):1–7, 2011.

\bibitem{kit/gem5-1}
Nathan Binkert, Bradford Beckmann, Gabriel Black, Steven~K. Reinhardt, Ali
  Saidi, Arkaprava Basu, Joel Hestness, Derek~R. Hower, Tushar Krishna, Somayeh
  Sardashti, Rathijit Sen, Korey Sewell, Muhammad Shoaib, Nilay Vaish, Mark~D.
  Hill, and David~A. Wood.
\newblock The gem5 simulator.
\newblock {\em SIGARCH Comput. Archit. News}, 39(2):1–7, aug 2011.

\bibitem{kit5}
Amirali Boroumand, Saugata Ghose, Youngsok Kim, Rachata Ausavarungnirun, Eric
  Shiu, Rahul Thakur, Daehyun Kim, Aki Kuusela, Allan Knies, Parthasarathy
  Ranganathan, et~al.
\newblock Google workloads for consumer devices: Mitigating data movement
  bottlenecks.
\newblock In {\em Proceedings of the Twenty-Third International Conference on
  Architectural Support for Programming Languages and Operating Systems}, pages
  316--331, 2018.

\bibitem{kit/dramScale1}
Kevin~K Chang.
\newblock {\em Understanding and improving the latency of DRAM-based memory
  systems}.
\newblock PhD thesis, Carnegie Mellon University, 2017.

\bibitem{kit/reram-tsmc4}
Meng-Fan Chang, Jui-Jen Wu, Tun-Fei Chien, Yen-Chen Liu, Ting-Chin Yang,
  Wen-Chao Shen, Ya-Chin King, Chorng-Jung Lin, Ku-Feng Lin, Yu-Der Chih,
  Sreedhar Natarajan, and Jonathan Chang.
\newblock 19.4 embedded 1mb reram in 28nm cmos with 0.27-to-1v read using
  swing-sample-and-couple sense amplifier and self-boost-write-termination
  scheme.
\newblock In {\em 2014 IEEE International Solid-State Circuits Conference
  Digest of Technical Papers (ISSCC)}, pages 332--333, 2014.

\bibitem{NEUROSIM}
Pai-Yu Chen, Xiaochen Peng, and Shimeng Yu.
\newblock Neurosim: A circuit-level macro model for benchmarking neuro-inspired
  architectures in online learning.
\newblock {\em IEEE Transactions on Computer-Aided Design of Integrated
  Circuits and Systems}, 37(12):3067--3080, 2018.

\bibitem{HeapWL}
Xianzhang Chen, Zhuge Qingfeng, Qiang Sun, Edwin H.-M. Sha, Shouzhen Gu,
  Chaoshu Yang, and Chun~Jason Xue.
\newblock A wear-leveling-aware fine-grained allocator for non-volatile memory.
\newblock In {\em Proceedings of the 56th Annual Design Automation Conference
  2019}, DAC '19, New York, NY, USA, 2019. Association for Computing Machinery.

\bibitem{kit/pcm-ibm3}
W.~C. Chien, L.~M. Gignac, Y.~C. Chou, C.~H. Yang, N.~Gong, H.~Y. Ho, C.~W.
  Yeh, H.~Y. Cheng, W.~Kim, I.~T. Kuo, E.~K. Lai, C.~W. Cheng, L.~Buzi, A.~Ray,
  C.~S. Hsu, D.~Daudelin, R.~L. Bruce, M.~BrightSky, and H.~L. Lung.
\newblock Device study on ots-pcm for persistent memory application :
  Ibm/macronix phase change memory joint project.
\newblock In {\em 2022 6th IEEE Electron Devices Technology \& Manufacturing
  Conference (EDTM)}, pages 327--329, 2022.

\bibitem{kit/stt-tsmc1}
Yu-Der Chih, Yi-Chun Shih, Chia-Fu Lee, Yen-An Chang, Po-Hao Lee, Hon-Jarn Lin,
  Yu-Lin Chen, Chieh-Pu Lo, Meng-Chun Shih, Kuei-Hung Shen, Harry Chuang, and
  Tsung-Yung~Jonathan Chang.
\newblock 13.3 a 22nm 32mb embedded stt-mram with 10ns read speed, 1m cycle
  write endurance, 10 years retention at 150°c and high immunity to magnetic
  field interference.
\newblock In {\em 2020 IEEE International Solid-State Circuits Conference -
  (ISSCC)}, pages 222--224, 2020.

\bibitem{10145822}
Jeongdong Choe.
\newblock Recent technology insights on stt-mram: Structure, materials, and
  process integration.
\newblock In {\em 2023 IEEE International Memory Workshop (IMW)}, pages 1--4,
  2023.

\bibitem{6176872}
Youngdon Choi, Ickhyun Song, Mu-Hui Park, Hoeju Chung, Sanghoan Chang,
  Beakhyoung Cho, Jinyoung Kim, Younghoon Oh, Duckmin Kwon, Jung Sunwoo, Junho
  Shin, Yoohwan Rho, Changsoo Lee, Min~Gu Kang, Jaeyun Lee, Yongjin Kwon,
  Soehee Kim, Jaehwan Kim, Yong-Jun Lee, Qi~Wang, Sooho Cha, Sujin Ahn, Hideki
  Horii, Jaewook Lee, Kisung Kim, Hansung Joo, Kwangjin Lee, Yeong-Taek Lee,
  Jeihwan Yoo, and Gitae Jeong.
\newblock A 20nm 1.8v 8gb pram with 40mb/s program bandwidth.
\newblock In {\em 2012 IEEE International Solid-State Circuits Conference},
  pages 46--48, 2012.

\bibitem{scons}
William Deegan.
\newblock {S}{C}ons: {A} software construction tool --- scons.org.
\newblock \url{https://scons.org/}.
\newblock [Accessed 13-01-2025].

\bibitem{devaux2019true}
Fabrice Devaux.
\newblock The true processing in memory accelerator.
\newblock In {\em 2019 IEEE Hot Chips 31 Symposium (HCS)}, pages 1--24. IEEE
  Computer Society, 2019.

\bibitem{6218223}
Xiangyu Dong, Cong Xu, Yuan Xie, and Norman~P. Jouppi.
\newblock Nvsim: A circuit-level performance, energy, and area model for
  emerging nonvolatile memory.
\newblock {\em IEEE Transactions on Computer-Aided Design of Integrated
  Circuits and Systems}, 31(7):994--1007, 2012.

\bibitem{ellis:2007}
Brian Ellis, Jeffrey Stylos, and Brad Myers.
\newblock The factory pattern in api design: A usability evaluation.
\newblock In {\em 29th International Conference on Software Engineering
  (ICSE'07)}, pages 302--312, 2007.

\bibitem{escuin_hpca23}
Carlos Escuin, Asif~Ali Khan, Pablo Ibáñez-Marín, Teresa Monreal, Jeronimo
  Castrillon, and Víctor Viñals-Yúfera.
\newblock Compression-aware and performance-efficient insertion policies for
  long-lasting hybrid llcs.
\newblock In {\em the 29th IEEE International Symposium on High-Performance
  Computer Architecture (HPCA'23)}, pages 179--192, Los Alamitos, CA, USA,
  March 2023. IEEE, IEEE Computer Society.

\bibitem{kit/ext4}
Anteneh Gebregiorgis, Hoang~Anh Du~Nguyen, Jintao Yu, Rajendra Bishnoi,
  Mottaqiallah Taouil, Francky Catthoor, and Said Hamdioui.
\newblock A survey on memory-centric computer architectures.
\newblock {\em ACM Journal on Emerging Technologies in Computing Systems
  (JETC)}, 18(4):1--50, 2022.

\bibitem{kit/ext3}
Saugata Ghose, Amirali Boroumand, Jeremie~S Kim, Juan G{\'o}mez-Luna, and Onur
  Mutlu.
\newblock Processing-in-memory: A workload-driven perspective.
\newblock {\em IBM Journal of Research and Development}, 63(6):3--1, 2019.

\bibitem{10750212}
Leandro~M. Giacomini~Rocha, Mohamed Naeim, Guilherme Paim, Moritz Brunion,
  Priya Venugopal, Dragomir Milojevic, James Myers, Mustafa Badaroglu, Marian
  Verlhest, Julien Ryckaert, and Dwaipayan Biswas.
\newblock System-technology co-optimization for dense edge architectures using
  3d integration and non-volatile memory.
\newblock {\em IEEE Journal on Exploratory Solid-State Computational Devices
  and Circuits}, pages 1--1, 2024.

\bibitem{guthaus:2001}
M.R. Guthaus, J.S. Ringenberg, D.~Ernst, T.M. Austin, T.~Mudge, and R.B. Brown.
\newblock Mibench: A free, commercially representative embedded benchmark
  suite.
\newblock In {\em IEEE International Workshop on Workload Characterization},
  pages 3--14, 2001.

\bibitem{kit/ext8}
Wilfried Haensch, Anand Raghunathan, Kaushik Roy, Bhaswar Chakrabarti,
  Charudatta~M Phatak, Cheng Wang, and Supratik Guha.
\newblock Compute in-memory with non-volatile elements for neural networks: A
  review from a co-design perspective.
\newblock {\em Advanced Materials}, 35(37):2204944, 2023.

\bibitem{hakert:2020:softwear}
Christian Hakert, Kuan-Hsun Chen, Paul~R. Genssler, Georg von~der Brüggen,
  Lars Bauer, Hussam Amrouch, Jian-Jia Chen, and Jörg Henkel.
\newblock Softwear: Software-only in-memory wear-leveling for non-volatile main
  memory, 2020.

\bibitem{hakert:2020:split}
Christian Hakert, Kuan-Hsun Chen, Simon Kuenzer, Sharan Santhanam, Shuo-Han
  Chen, Yuan-Hao Chang, Felipe Huici, and Jian-Jia Chen.
\newblock Split'n trace nvm: Leveraging library oses for semantic memory
  tracing.
\newblock In {\em 2020 9th Non-Volatile Memory Systems and Applications
  Symposium (NVMSA)}, pages 1--6, 2020.

\bibitem{hakert:2022}
Christian Hakert, Kuan-Hsun Chen, Horst Schirmeier, Lars Bauer, Paul~R.
  Genssler, Georg von~der Br\"{u}ggen, Hussam Amrouch, J\"{o}rg Henkel, and
  Jian-Jia Chen.
\newblock Software-managed read and write wear-leveling for non-volatile main
  memory.
\newblock {\em ACM Trans. Embed. Comput. Syst.}, 21(1), February 2022.

\bibitem{hakert:2020:base}
Christian Hakert, Kuan-Hsun Chen, Mikail Yayla, Georg von~der Brüggen,
  Sebastian Blömeke, and Jian-Jia Chen.
\newblock Software-based memory analysis environments for in-memory
  wear-leveling.
\newblock In {\em 2020 25th Asia and South Pacific Design Automation Conference
  (ASP-DAC)}, pages 651--658, 2020.

\bibitem{hameed_tetc21}
Fazal Hameed, Asif~Ali Khan, and Jeronimo Castrillon.
\newblock {ALPHA}: A novel algorithm-hardware co-design for accelerating {DNA}
  seed location filtering.
\newblock page 12 pp., June 2021.

\bibitem{kit2}
John~L. Hennessy and David~A. Patterson.
\newblock A new golden age for computer architecture.
\newblock {\em Commun. ACM}, 62(2):48–60, jan 2019.

\bibitem{hoelscher:2022}
Nils Hölscher, Christian Hakert, Hassan Nassar, Kuan-Hsun Chen, Lars Bauer,
  Jian-Jia Chen, and Jörg Henkel.
\newblock Memory carousel: Llvm-based bitwise wear-leveling for non-volatile
  main memory.
\newblock {\em IEEE Transactions on Computer-Aided Design of Integrated
  Circuits and Systems}, pages 1--1, 2022.

\bibitem{hoelscher:2023}
Nils Hölscher, Minh~Duy Truong, Christian Hakert, Tristan Seidl, Kuan-Hsun
  Chen, and Jian-Jia Chen.
\newblock Rapid nvm simulation and analysis on single bit granularity featuring
  gem5 and nvmain.
\newblock In {\em 2023 IEEE 12th Non-Volatile Memory Systems and Applications
  Symposium (NVMSA)}, pages 50--55, 2023.

\bibitem{7479181}
Mohsen Imani, Shruti Patil, and Tajana Rosing.
\newblock Low power data-aware stt-ram based hybrid cache architecture.
\newblock In {\em 2016 17th International Symposium on Quality Electronic
  Design (ISQED)}, pages 88--94, 2016.

\bibitem{kit/hbm}
Hongshin Jun, Jinhee Cho, Kangseol Lee, Ho-Young Son, Kwiwook Kim, Hanho Jin,
  and Keith Kim.
\newblock Hbm (high bandwidth memory) dram technology and architecture.
\newblock In {\em 2017 IEEE International Memory Workshop (IMW)}, pages 1--4.
  IEEE, 2017.

\bibitem{kit3}
Gokcen Kestor, Roberto Gioiosa, Darren~J Kerbyson, and Adolfy Hoisie.
\newblock Quantifying the energy cost of data movement in scientific
  applications.
\newblock In {\em 2013 IEEE international symposium on workload
  characterization (IISWC)}, pages 56--65. IEEE, 2013.

\bibitem{khan2019rtsim}
Asif~Ali Khan, Fazal Hameed, Robin Bl{\"a}sing, Stuart Parkin, and Jeronimo
  Castrillon.
\newblock Rtsim: A cycle-accurate simulator for racetrack memories.
\newblock {\em IEEE Computer Architecture Letters}, 18(1):43--46, 2019.

\bibitem{asif:18}
Asif~Ali Khan, Fazal Hameed, and Jeronimo Castrillon.
\newblock Nvmain extension for multi-level cache systems.
\newblock In {\em Proceedings of the Rapido'18 Workshop on Rapid Simulation and
  Performance Evaluation: Methods and Tools}, RAPIDO '18, New York, NY, USA,
  2018. Association for Computing Machinery.

\bibitem{khan_cimlandscape_2024}
Asif~Ali Khan, João Paulo C.~De Lima, Hamid Farzaneh, and Jeronimo Castrillon.
\newblock The landscape of compute-near-memory and compute-in-memory: A
  research and commercial overview, January 2024.

\bibitem{khan2023downshift}
Asif~Ali Khan, Sebastien Ollivier, Fazal Hameed, Jeronimo Castrillon, and
  Alex~K Jones.
\newblock Downshift: Tuning shift reduction with reliability for racetrack
  memories.
\newblock {\em IEEE Transactions on Computers}, 72(9):2585--2599, 2023.

\bibitem{kuenzer:2021}
Simon Kuenzer, Vlad-Andrei B\u{a}doiu, Hugo Lefeuvre, Sharan Santhanam,
  Alexander Jung, Gaulthier Gain, Cyril Soldani, Costin Lupu, \c{S}tefan
  Teodorescu, Costi R\u{a}ducanu, Cristian Banu, Laurent Mathy, R\u{a}zvan
  Deaconescu, Costin Raiciu, and Felipe Huici.
\newblock Unikraft: Fast, specialized unikernels the easy way.
\newblock In {\em 16th European Conference on Computer Systems}, page
  376–394, 2021.

\bibitem{dwm-nvmain:19}
Jinzhi Lai, Jueping Cai, Lai Liu, and Zhuoye Huang.
\newblock A design and analysis perspective on architecting memory using
  domain-wall memory.
\newblock In {\em 2019 IEEE International Conference on Smart Internet of
  Things (SmartIoT)}, pages 454--458, 2019.

\bibitem{1555815}
Benjamin~C. Lee, Engin Ipek, Onur Mutlu, and Doug Burger.
\newblock Architecting phase change memory as a scalable dram alternative.
\newblock {\em SIGARCH Comput. Archit. News}, 37(3):2–13, June 2009.

\bibitem{kit/stt-samsung2}
K.~Lee, D.~S. Kim, J.~H. Bak, S.~P. Ko, W.~C. Lim, H.~C. Shin, J.~H. Lee, J.~H.
  Park, J.~H. Jeong, J.~M. Lee, T.~Kai, H.~Sato, J.~W. Lee, K.~H. Ryu, Y.~J.
  Kim, S.~H. Han, B.~Y. Seo, K.~S. Suh, H.~H. Kim, H.~T. Jung, D.~H. Jang,
  N.~Y. Ji, M.~J. Eom, I.~H. Kim, K.~Lee, K.~H. Hwang, Y.~J. Song, and H.~S.
  Kim.
\newblock 28nm cis-compatible embedded stt-mram for frame buffer memory.
\newblock In {\em 2021 IEEE International Electron Devices Meeting (IEDM)},
  pages 2.1.1--2.1.4, 2021.

\bibitem{Stack}
Qingan Li, Yanxiang He, Yong Chen, Chun~Jason Xue, Nan Jiang, and Chao Xu.
\newblock A wear-leveling-aware dynamic stack for pcm memory in embedded
  systems.
\newblock pages 1--4, 2014.

\bibitem{HeapStackWL}
Wei Li, Ziqi Shuai, Chun~Jason Xue, Mengting Yuan, and Qingan Li.
\newblock A wear leveling aware memory allocator for both stack and heap
  management in pcm-based main memory systems.
\newblock In {\em 2019 Design, Automation and Test in Europe Conference and
  Exhibition (DATE)}, pages 228--233, 2019.

\bibitem{kit/db2}
Yinan Li and Jignesh~M Patel.
\newblock Bitweaving: Fast scans for main memory data processing.
\newblock In {\em Proceedings of the 2013 ACM SIGMOD International Conference
  on Management of Data}, pages 289--300, 2013.

\bibitem{kit/app6}
Yinan Li and Jignesh~M Patel.
\newblock Bitweaving: Fast scans for main memory data processing.
\newblock In {\em Proceedings of the 2013 ACM SIGMOD International Conference
  on Management of Data}, pages 289--300, 2013.

\bibitem{kit/dramScale2}
Kevin Lim, Jichuan Chang, Trevor Mudge, Parthasarathy Ranganathan, Steven~K
  Reinhardt, and Thomas~F Wenisch.
\newblock Disaggregated memory for expansion and sharing in blade servers.
\newblock {\em ACM SIGARCH computer architecture news}, 37(3):267--278, 2009.

\bibitem{kit/dramScale3}
Jamie Liu, Ben Jaiyen, Yoongu Kim, Chris Wilkerson, and Onur Mutlu.
\newblock An experimental study of data retention behavior in modern dram
  devices: Implications for retention time profiling mechanisms.
\newblock {\em ACM SIGARCH Computer Architecture News}, 41(3):60--71, 2013.

\bibitem{kit/img1}
William~E. Lorensen and Harvey~E. Cline.
\newblock Marching cubes: A high resolution 3d surface construction algorithm.
\newblock {\em SIGGRAPH Comput. Graph.}, 21(4):163–169, August 1987.

\bibitem{kit/app4}
William~E Lorensen and Harvey~E Cline.
\newblock Marching cubes: A high resolution 3d surface construction algorithm.
\newblock In {\em Seminal graphics: pioneering efforts that shaped the field},
  pages 347--353. 1998.

\bibitem{kit/gem5-2}
Jason Lowe-Power, Abdul~Mutaal Ahmad, Ayaz Akram, Mohammad Alian, Rico
  Amslinger, Matteo Andreozzi, Adrià Armejach, Nils Asmussen, Brad Beckmann,
  Srikant Bharadwaj, Gabe Black, Gedare Bloom, Bobby~R. Bruce, Daniel~Rodrigues
  Carvalho, Jeronimo Castrillon, Lizhong Chen, Nicolas Derumigny, Stephan
  Diestelhorst, Wendy Elsasser, Carlos Escuin, Marjan Fariborz, Amin
  Farmahini-Farahani, Pouya Fotouhi, Ryan Gambord, Jayneel Gandhi, Dibakar
  Gope, Thomas Grass, Anthony Gutierrez, Bagus Hanindhito, Andreas Hansson,
  Swapnil Haria, Austin Harris, Timothy Hayes, Adrian Herrera, Matthew
  Horsnell, Syed Ali~Raza Jafri, Radhika Jagtap, Hanhwi Jang, Reiley Jeyapaul,
  Timothy~M. Jones, Matthias Jung, Subash Kannoth, Hamidreza Khaleghzadeh,
  Yuetsu Kodama, Tushar Krishna, Tommaso Marinelli, Christian Menard, Andrea
  Mondelli, Miquel Moreto, Tiago Mück, Omar Naji, Krishnendra Nathella, Hoa
  Nguyen, Nikos Nikoleris, Lena~E. Olson, Marc Orr, Binh Pham, Pablo Prieto,
  Trivikram Reddy, Alec Roelke, Mahyar Samani, Andreas Sandberg, Javier
  Setoain, Boris Shingarov, Matthew~D. Sinclair, Tuan Ta, Rahul Thakur, Giacomo
  Travaglini, Michael Upton, Nilay Vaish, Ilias Vougioukas, William Wang,
  Zhengrong Wang, Norbert Wehn, Christian Weis, David~A. Wood, Hongil Yoon, and
  Éder F.~Zulian.
\newblock The gem5 simulator: Version 20.0+, 2020.

\bibitem{8716297}
Jing-Yuan Luo, Hsiang-Yun Cheng, Ing-Chao Lin, and Da-Wei Chang.
\newblock Tap: Reducing the energy of asymmetric hybrid last-level cache via
  thrashing aware placement and migration.
\newblock {\em IEEE Transactions on Computers}, 68(12):1704--1719, 2019.

\bibitem{stream_bench}
J.~D. McCalpin.
\newblock Memory bandwidth and machine balance in current high performance
  computers.
\newblock {\em IEEE Technical Committee on Computer Architecture (TCCA)
  Newsletter}, Dec 1995.

\bibitem{kit/ext7}
Dejan Milojicic, Kirk Bresniker, Gary Campbell, Paolo Faraboschi, John~Paul
  Strachan, and Stan Williams.
\newblock Computing in-memory, revisited.
\newblock In {\em 2018 IEEE 38th International Conference on Distributed
  Computing Systems (ICDCS)}, pages 1300--1309. IEEE, 2018.

\bibitem{thermal_soc:24}
Sharli Mishra, Bas Vermeersch, Sankatali Venkateswarlu, H.~Kukner,
  G.~Mirabelli, F.~Bufler, M.~Brunion, Dawit Abdi, Herman Oprins, D.~Biswas,
  O.~Zografos, F.~Catthoor, P.~Weckx, G.~Hellings, J.~Myers, and J.~Ryckaert.
\newblock Thermal considerations for block-level ppa assessment in angstrom
  era: A comparison study of nanosheet fets (a10) \& complementary fets (a5).
\newblock pages 1--2, 06 2024.

\bibitem{CacheWL}
Sparsh Mittal, Jeffrey~S. Vetter, and Dong Li.
\newblock Writesmoothing: Improving lifetime of non-volatile caches using
  intra-set wear-leveling.
\newblock In {\em Proceedings of the 24th Edition of the Great Lakes Symposium
  on VLSI}, GLSVLSI '14, page 139–144, New York, NY, USA, 2014. Association
  for Computing Machinery.

\bibitem{kit/emb}
Cleve Moler.
\newblock Matrix computation on distributed memory multiprocessors.
\newblock {\em Hypercube Multiprocessors}, 86(181-195):31, 1986.

\bibitem{PiMulator}
Sergiu Mosanu, Mohammad~Nazmus Sakib, Tommy Tracy, Ersin Cukurtas, Alif Ahmed,
  Preslav Ivanov, Samira Khan, Kevin Skadron, and Mircea Stan.
\newblock Pimulator: a fast and flexible processing-in-memory emulation
  platform.
\newblock In {\em 2022 Design, Automation \& Test in Europe Conference \&
  Exhibition (DATE)}, pages 1473--1478, 2022.

\bibitem{kit/ext1}
Onur Mutlu, Saugata Ghose, Juan G{\'o}mez-Luna, and Rachata Ausavarungnirun.
\newblock Processing data where it makes sense: Enabling in-memory computation.
\newblock {\em Microprocessors and Microsystems}, 67:28--41, 2019.

\bibitem{kit/ext2}
Onur Mutlu, Saugata Ghose, Juan G{\'o}mez-Luna, and Rachata Ausavarungnirun.
\newblock A modern primer on processing in memory.
\newblock In {\em Emerging computing: from devices to systems: looking beyond
  Moore and Von Neumann}, pages 171--243. Springer, 2022.

\bibitem{kit/hmc}
{Nuvation Engineering}.
\newblock Hmc 30g vsr specification.
\newblock Nuvation Engineering, 2024.
\newblock Accessed: 2024-10-17.

\bibitem{PageWL}
Chen Pan, Shouzhen Gu, Mimi Xie, Yongpan Liu, Chun~Jason Xue, and Jingtong Hu.
\newblock Wear-leveling aware page management for non-volatile main memory on
  embedded systems.
\newblock volume~2, pages 129--142, 2016.

\bibitem{kit4}
Dhinakaran Pandiyan and Carole-Jean Wu.
\newblock Quantifying the energy cost of data movement for emerging smart phone
  workloads on mobile platforms.
\newblock In {\em 2014 IEEE International Symposium on Workload
  Characterization (IISWC)}, pages 171--180. IEEE, 2014.

\bibitem{9773239}
Lillian Pentecost, Alexander Hankin, Marco Donato, Mark Hempstead, Gu-Yeon Wei,
  and David Brooks.
\newblock Nvmexplorer: A framework for cross-stack comparisons of embedded
  non-volatile memories.
\newblock In {\em 2022 IEEE International Symposium on High-Performance
  Computer Architecture (HPCA)}, pages 938--956, 2022.

\bibitem{philofsky:1996}
E.M. Philofsky.
\newblock Fram-the ultimate memory.
\newblock In {\em Proceedings of Nonvolatile Memory Technology Conference},
  pages 99--104, 1996.

\bibitem{poremba:2015}
Matthew Poremba, Tao Zhang, and Yuan Xie.
\newblock {NVMain 2.0}: A user-friendly memory simulator to model
  (non-)volatile memory systems.
\newblock {\em IEEE Computer Architecture Letters}, 14(2):140--143, 2015.

\bibitem{rram:22}
Kartik Prabhu, Albert Gural, Zainab~F. Khan, Robert~M. Radway, Massimo
  Giordano, Kalhan Koul, Rohan Doshi, John~W. Kustin, Timothy Liu, Gregorio~B.
  Lopes, Victor Turbiner, Win-San Khwa, Yu-Der Chih, Meng-Fan Chang, Guénolé
  Lallement, Boris Murmann, Subhasish Mitra, and Priyanka Raina.
\newblock Chimera: A 0.92-tops, 2.2-tops/w edge ai accelerator with 2-mbyte
  on-chip foundry resistive ram for efficient training and inference.
\newblock {\em IEEE Journal of Solid-State Circuits}, 57(4):1013--1026, 2022.

\bibitem{5416645}
Moinuddin~K. Qureshi, Michele~M. Franceschini, and Luis~A. Lastras-Montaño.
\newblock Improving read performance of phase change memories via write
  cancellation and write pausing.
\newblock In {\em HPCA - 16 2010 The Sixteenth International Symposium on
  High-Performance Computer Architecture}, pages 1--11, 2010.

\bibitem{rogers:1995}
Anne Rogers, Martin~C. Carlisle, John~H. Reppy, and Laurie~J. Hendren.
\newblock Supporting dynamic data structures on distributed-memory machines.
\newblock {\em ACM Trans. Program. Lang. Syst.}, 17(2):233–263, mar 1995.

\bibitem{Sim2PIM}
Paulo~C. Santos, Bruno~E. Forlin, and Luigi Carro.
\newblock Sim2pim: A fast method for simulating host independent \& pim
  agnostic designs.
\newblock In {\em 2021 Design, Automation \& Test in Europe Conference \&
  Exhibition (DATE)}, pages 226--231, 2021.

\bibitem{9399236}
Arindam Sarkar, Newton Singh, Varun Venkitaraman, and Virendra Singh.
\newblock Dam: Deadblock aware migration techniques for stt-ram-based hybrid
  caches.
\newblock {\em IEEE Computer Architecture Letters}, 20(1):62--4, 2021.

\bibitem{kit/rowclone}
Vivek Seshadri, Yoongu Kim, Chris Fallin, Donghyuk Lee, Rachata
  Ausavarungnirun, Gennady Pekhimenko, Yixin Luo, Onur Mutlu, Phillip~B
  Gibbons, Michael~A Kozuch, et~al.
\newblock Rowclone: Fast and energy-efficient in-dram bulk data copy and
  initialization.
\newblock In {\em Proceedings of the 46th Annual IEEE/ACM International
  Symposium on Microarchitecture}, pages 185--197, 2013.

\bibitem{ambit}
Vivek Seshadri, Donghyuk Lee, Thomas Mullins, Hasan Hassan, Amirali Boroumand,
  Jeremie Kim, Michael~A Kozuch, Onur Mutlu, Phillip~B Gibbons, and Todd~C
  Mowry.
\newblock Ambit: In-memory accelerator for bulk bitwise operations using
  commodity dram technology.
\newblock In {\em Proceedings of the 50th Annual IEEE/ACM International
  Symposium on Microarchitecture}, pages 273--287, 2017.

\bibitem{kit/img2}
Frank~Y Shih.
\newblock {\em Image processing and mathematical morphology: fundamentals and
  applications}.
\newblock CRC press, 2017.

\bibitem{kit/app3}
Frank~Y Shih.
\newblock {\em Image processing and mathematical morphology: fundamentals and
  applications}.
\newblock CRC press, 2017.

\bibitem{fram:24}
Mandeep Singh, Tarun Chaudhary, Balwinder Raj, Suman Tripathi, and k.srinavso
  Rao.
\newblock {\em Integrated Devices for Artificial Intelligence and VLSI}.
\newblock 09 2024.

\bibitem{rowMirroring}
Simranjeet Singh, Ankit Bende, Chandan~Kumar Jha, Vikas Rana, Rolf Drechsler,
  Sachin Patkar, and Farhad Merchant.
\newblock In-memory mirroring: Cloning without reading.
\newblock {\em arXiv preprint arXiv:2407.02921}, 2024.

\bibitem{rram:23}
T.~Srimani, A.~Bechdolt, S.~Choi, C.~Gilardi, A.~Kasperovich, S.~Li, Q.~Lin,
  M.~Malakoutian, P.~McEwen, R.M. Radway, D.~Rich, A.C. Yu, S.~Fuller,
  S.~Achour, S.~Chowdhury, H.-S.~P. Wong, M.~Shulaker, and S.~Mitra.
\newblock N3xt 3d technology foundations and their lab-to-fab: Omni 3d logic,
  logic+memory ultra-dense 3d, 3d thermal scaffolding.
\newblock In {\em 2023 International Electron Devices Meeting (IEDM)}, pages
  1--4, 2023.

\bibitem{kit/db1}
Kurt Stockinger, John Cieslewicz, Kesheng Wu, Doron Rotem, and Arie Shoshani.
\newblock Using bitmap index for joint queries on structured and text data.
\newblock In {\em New Trends in Data Warehousing and Data Analysis}, pages
  1--23. Springer, 2008.

\bibitem{kit/app1}
Kurt Stockinger, John Cieslewicz, Kesheng Wu, Doron Rotem, and Arie Shoshani.
\newblock Using bitmap index for joint queries on structured and text data.
\newblock In {\em New Trends in Data Warehousing and Data Analysis}, pages
  1--23. Springer, 2008.

\bibitem{4798259}
Guangyu Sun, Xiangyu Dong, Yuan Xie, Jian Li, and Yiran Chen.
\newblock A novel architecture of the 3d stacked mram l2 cache for cmps.
\newblock In {\em 2009 IEEE 15th International Symposium on High Performance
  Computer Architecture}, pages 239--249, 2009.

\bibitem{kit1}
J.~von Neumann.
\newblock First draft of a report on the edvac.
\newblock {\em IEEE Annals of the History of Computing}, 15(4):27--75, 1993.

\bibitem{Wilbert:2024a}
Nils Wilbert, Stefan Wildermann, and J\"{u}rgen Teich.
\newblock Hybrid cache design under varying power supply stability - a
  comparative study.
\newblock In {\em Proceedings of the 10th International Symposium on Memory
  Systems}, MEMSYS \'24. Association for Computing Machinery, 2024.

\bibitem{Wilbert:24b}
Nils Wilbert, Stefan Wildermann, and J\"{u}rgen Teich.
\newblock To keep or not to keep - the volatility of replacement policy
  metadata in hybrid caches.
\newblock In {\em Proceedings of the 2nd Workshop on Disruptive Memory
  Systems}, DIMES '24, page 17–24, New York, NY, USA, 2024. Association for
  Computing Machinery.

\bibitem{Loop2RecWL}
Jifeng Wu, Wei Li, Libing Wu, Mengting Yuan, Chun~Jason Xue, Jingling Xue, and
  Qingan Li.
\newblock Effective stack wear leveling for nvm.
\newblock {\em IEEE Transactions on Computer-Aided Design of Integrated
  Circuits and Systems}, pages 1--1, 2023.

\bibitem{pcm:2024}
Xiangjin Wu, Asir Intisar~Khan, Hengyuan Lee, Chen-Feng Hsu, Huairuo Zhang,
  Heshan Yu, Neel Roy, Albert Davydov, I.~Takeuchi, Xinyu Bao, H-S Wong, and
  Eric Pop.
\newblock Novel nanocomposite-superlattices for low energy and high stability
  nanoscale phase- change memory.
\newblock {\em Nature Communications}, 15, 01 2024.

\bibitem{5090762}
Xiaoxia Wu, Jian Li, Lixin Zhang, Evan Speight, and Yuan Xie.
\newblock Power and performance of read-write aware hybrid caches with
  non-volatile memories.
\newblock In {\em 2009 Design, Automation \& Test in Europe Conference \&
  Exhibition}, pages 737--742, 2009.

\bibitem{MNSIM}
Lixue Xia, Boxun Li, Tianqi Tang, Peng Gu, Xiling Yin, Wenqin Huangfu, Pai-Yu
  Chen, Shimeng Yu, Yu~Cao, Yu~Wang, Yuan Xie, and Huazhong Yang.
\newblock Mnsim: Simulation platform for memristor-based neuromorphic computing
  system.
\newblock In {\em 2016 Design, Automation \& Test in Europe Conference \&
  Exhibition (DATE)}, pages 469--474, 2016.

\bibitem{kit/dna1}
Hongyi Xin, John Greth, John Emmons, Gennady Pekhimenko, Carl Kingsford, Can
  Alkan, and Onur Mutlu.
\newblock {Shifted Hamming distance: a fast and accurate SIMD-friendly filter
  to accelerate alignment verification in read mapping}.
\newblock {\em Bioinformatics}, 31(10):1553--1560, 01 2015.

\bibitem{kit/app5}
Hongyi Xin, John Greth, John Emmons, Gennady Pekhimenko, Carl Kingsford, Can
  Alkan, and Onur Mutlu.
\newblock {Shifted Hamming distance: a fast and accurate SIMD-friendly filter
  to accelerate alignment verification in read mapping}.
\newblock {\em Bioinformatics}, 31(10):1553--1560, 01 2015.

\bibitem{PIMSim}
Sheng Xu, Xiaoming Chen, Ying Wang, Yinhe Han, Xuehai Qian, and Xiaowei Li.
\newblock Pimsim: A flexible and detailed processing-in-memory simulator.
\newblock {\em IEEE Computer Architecture Letters}, 18(1):6--9, 2019.

\bibitem{yang:2007}
Byung-Do Yang, Jae-Eun Lee, Jang-Su Kim, Junghyun Cho, Seung-Yun Lee, and
  Byoung-Gon Yu.
\newblock A low power phase-change random access memory using a data-comparison
  write scheme.
\newblock In {\em 2007 IEEE International Symposium on Circuits and Systems
  (ISCAS)}, pages 3014--3017, 2007.

\bibitem{MultiPIM}
Chao Yu, Sihang Liu, and Samira Khan.
\newblock Multipim: A detailed and configurable multi-stack
  processing-in-memory simulator.
\newblock {\em IEEE Computer Architecture Letters}, 20(1):54--57, 2021.

\end{thebibliography}

\section*{APPENDICES}\label{sec:appendices}
\subsection*{Toolchain Setup}
\label{app:setup}
In this appendix, we describe how to set up the toolchain described in this work. Toolchain's respository on Github~\cite{repo} also includes a README-file with more details on the setup instructions. This setup is based on our previous work~\cite{hoelscher:2023}, in which we made the toolchain more accessible by integrating it with an IDE.

To work with the presented toolchain, we recommend setting up the environment
the same way we do. For this, Visual Studio Code~\cite{vscode}, its extensions for working with remote
development container~\cite{vscodeextensions} as well as Docker~\cite{docker} are required. Please see their respective documentation for setup instructions or the repository's README-file. Alternatively, you can use your own environment, as long as Docker is supported.

To begin with, go ahead and download the repository. After downloading, go into the repository's root directory and execute~\cref{lst:submodule}
\begin{lstlisting}[caption={Submodule initialization},label=lst:submodule,language=bash]
$ git submodule init
$ git submodule update
\end{lstlisting}
This will set up the source code for gem5, NVMain, Unikraft, and the benchmarks.

Now start Visual Studio Code and open up the repository's directory. A prompt should pop up,
asking, if you want to reopen the directory in a development container. Go ahead and do so. If not asked, the directory can be reopened in a container via the blue "Remote Window" icon in the bottom left of
Visual Studio Code.

Before using the toolchain, it needs to be compiled. For this, we have provided an example script
(\texttt{ExampleBuild.sh}) within the repository that you can use. It provides some general
commands that enable building and executing the toolchain. In its current version, it compiles the
toolchain with our bit flip simulation extension enabled (see section~\cref{subsec:tudo}). More
information on the compiled simulation environment can be found within the example script. Please
modify it to your liking or use your own build commands. However, in the context of this appendix, move into
the repository's root directory and run~\cref{lst:build}
\begin{lstlisting}[caption={Build gem5 and NVMain},label=lst:build,language=bash]
$ ./ExampleBuild fb
\end{lstlisting}
The command builds gem5 and NVMain. If at any time, you want to use the provided benchmark
applications, you first need to build them with~\cref{lst:benchmarks}
\begin{lstlisting}[caption={Build benchmarks},label=lst:benchmarks,language=bash]
$ ./ExampleBuild ba BENCHMARK_NAME
\end{lstlisting}

To start a simulation run~\cref{lst:simulation} with any application, e.g., one of the benchmarks.
\begin{lstlisting}[caption={Run simulation},label=lst:simulation,language=bash]
$ ./ExampleBuild e APPLICATION_NAME
\end{lstlisting}

Following these instructions puts the repository in a usable state that has our bit-flip simulation
extension enabled, i.e., you can directly start to follow section~\cref{subsec:tudo}. In case you
want to enable other extensions, please provide the respective flags when building gem5 and NVMain.
For reference, see~\cref{lst:noflags} compared to~\cref{lst:flags}.
\begin{lstlisting}[caption={Build without flags},label=lst:noflags,language=bash]
$ python3 `which scons` -j 8 EXTRAS=../nvmain ./build/ARM/gem5.fast
\end{lstlisting}
\begin{lstlisting}[caption={Build with the \texttt{CDNCcim} flag for X86 architecture, without the NVMain extension},label=lst:flags,language=bash]
$ python3 `which scons` CDNCcim=1 -j 8 EXTRAS=../nvmain ./build/X86/gem5.fast
\end{lstlisting}
We currently provide the following flags for the features we have extended the toolchain with:
\begin{itemize}
    \item \texttt{tu\_dortmund}: Bit flip simulation
	\item \texttt{CDNCcim}: Compute in Memory for NVM
	\item \texttt{hybrid\_cache}: Heterogeneous cache extension
\end{itemize}

While building (see~\cref{lst:noflags}), you may notice that the target binary is \texttt{gem5.fast} in the directory \texttt{ARM}. Specifying the binary type (\texttt{fast}, \texttt{debug}, \texttt{opt}) works solely by defining the target binary. For example, to include debug information, use \texttt{gem5.debug} instead. The same applies to the ISA, e.g., switching from ARM to X86, for both of which the toolchain has been tested~\footnote{
Due to an addressing issue when integrating NVMain with the latest gem5 version, the flag \texttt{X86} needs to be provided when targeting X86 architectures. With ARM, the flag can be ignored. Other ISAs are currently not handled.
}. For details on build versions, refer to gem5's official documentation~\cite{gem5:documentation}. The example script builds the \texttt{.fast}~-~version for ARM as shown in~\cref{lst:noflags}.
You might also need to build the binary for X86 target and m5 library, use the following command:
\begin{lstlisting}[caption={Build X86 target and m5 library},label=lst:gem5_x86,language=bash]
$ python3 `which scons` -j 8 EXTRAS=../nvmain ./build/X86/gem5.fast
$ python3 `which scons` -j 8 EXTRAS=gem5/util/nvmain -C gem5/util/m5 build/x86/out/m5
\end{lstlisting}

From this point, you can follow the case studies presented in this work. They all assume that the repository was set up using the instructions above. 
If a different repository state is required, the case study will provide the necessary steps. 
For more details on the overall toolchain setup, we refer the readers to the repository's README file and its components' respective documentation.
 
\subsection*{Main memory simulation, hybrid setup and new operations}
\label{app:tudr:main_memory}
Once you have set up the \textit{NVM\_Simulation} repository, switch to a different branch using the following command:
\begin{lstlisting}[caption={Build gem5 and m5 library},language=bash]
$ git checkout Tutorial-NVM
\end{lstlisting}
Then,  compile the \texttt{gem5.fast} binary for the X86 architecture by running the commands in Listing ~\cref{lst:gem5_x86}. After this, you are ready to proceed to run the different experiments.

\subsubsection*{Timing and scheduling policies}
\label{app:tudr:scheduling}
The first task runs a matrix multiplication in gem5. Compile the sample \texttt{mm.c} using the following command:
\begin{lstlisting}[language=bash]
$ gcc -O3 -static NVM_tutorial/mm.c -o NVM_tutorial/mm.bin
\end{lstlisting}

To obtain statistics for a specific code region, i.e., excluding the program's initialization or finalization phases and only focusing on the matmul computation kernel, the following m5 calls were be placed around the \emph{region of interest} (RoI):   

\begin{lstlisting}[language=C]
#include <gem5/m5ops.h>
...
m5_reset_stats(0, 0);
/*** Region of Interest ***/
m5_dump_stats(0, 0);
... 
\end{lstlisting}
This also requires the compile command to be updated to include the m5 library, as shown below:
\begin{lstlisting}[language=bash]
$ gcc -O3 -static NVM_tutorial/mm_m5.c -I gem5/include/ -lm5 -Lgem5/util/m5/build/x86/out -o NVM_tutorial/mm_m5.bin
\end{lstlisting}

To run this application in gem5 using NVM memory, use the following command, as mentioned in ~\cref{subsec:nvm-dram-methods}:

\begin{lstlisting}[language=bash]
$ gem5/build/X86/gem5.fast --outdir=FCFS example/se.py  --cpu-type=DerivO3CPU --caches --mem-type=NVMainMemory --nvmain- config=nvmain/Config/PCM_ISSCC_2012_4GB.config --cmd NVM_tutorial/mm_m5.bin &> FCFS/nvmain_stats.txt
\end{lstlisting}

With this command line, Gem5 records the statistics in the output directory (i.e., FCFS/stats.txt) for the first chunk of the RoI between the markers "Begin" and "End". The NVMain output is redirected (FCFS/nvmain\_stats.txt) and is divided into two chunks, starting with \texttt{i0.defaultMemory} and \texttt{i1.defaultMemory...}, where the first chunk corresponds to the RoI.  

As described in \cref{subsec:nvm-dram-methods}, you can change the memory controller in the NVMain configuration file to \texttt{FCFS, FRFCFS, FRFCFS-WQF} and re-run the above command to notice the different on the result (in the generated stats file). 
 
\subsubsection*{Hybrid main memory simulation}
\label{app:tudr:hybrid}
The previous task uses gem5+NVMain to simulate an NVM memory. To simulate a hybrid memory, open the gem5 system configuration file using:
\begin{lstlisting}[language=bash]
$ cd sims/tudr/nv-gem5
$ nano configs/deprecated/example/hybrid_example.py
\end{lstlisting}
In this file, one can configure the memory types and specify the number of channels for the system.
\begin{lstlisting}[language=Python]
### hybrid_example.py (Line 130)
args.mem_channels = 1 # single mem_ctrl
args.external_memory_system = 0
args.hybrid_channel = True
args.mem_type = "DDR4_2400_16x4"
args.nvm_type = "NVM_2400_1x64"
\end{lstlisting}

\begin{lstlisting}[language=Python]
### configs/common/MemConfig.py (Line 246)
# mem_ctrl.dram = dram_intf
### configs/common/MemConfig.py (Line 267)
mem_ctrls.append(nvm_intf.controller())

### hybrid_example.py (Line 130)
args.hybrid_channel = False # for separate mem_ctrls ...
system.mem_ctrls[0].dram.addr_mapping = "RoRaBaCoCh"
system.mem_ctrls[1].dram.addr_mapping = "RoRaBaCoCh"
\end{lstlisting}

To test this configuration, use the following commands to run the STREAM benchmark~\cite{stream_bench} on the gem5 simulator:

\begin{lstlisting}[language=bash]
$ gem5/build/X86/gem5.fast gem5/configs/deprecated/example/hybrid_example.py NVM_tutorial/stream_8k.bin
$ gem5/build/X86/gem5.fast gem5/configs/deprecated/example/hybrid_example.py NVM_tutorial/stream_8k_P.bin
\end{lstlisting}

\subsubsection*{Trace-based simulation}
\label{app:tudr:traceSim}
To view the NVMain trace format and a sample trace, run the following command:
\begin{lstlisting}[language=bash]
$ cat nvmain/Tests/Traces/hello_world.nvt
\end{lstlisting}

Use the following command to compile NVMain and simulate this trace file. 
\begin{lstlisting}[language=bash]
$ cd nvmain
$ python3 `which scons` --build-type=fast
$ ./nvmain.fast Config/PCM_ISSCC_2012_4GB.config Tests/Traces/hello_world.nvt 1000000
\end{lstlisting}

\subsubsection*{Adding new operations}
\label{app:tudr:adding_ops}
Use the following commands to test a trace file with rowclone requests.
\begin{lstlisting}[language=bash]
$ ./nvmain.fast Config/Config/2D_DRAM_example.config Tests/Traces/row_clone.nvt 1000
\end{lstlisting}
You can also view the trace file \texttt{Tests/Traces/row\_clone.nvt} to see the rowclone requests.  

\begin{lstlisting}[language=bash]
$ cat nvmain/Tests/Traces/row_clone.nvt
\end{lstlisting}

In the generated output, similar to the \texttt{read, write} requests statistics, you can also see the \texttt{rowclone} statistics. 

\subsection*{Access Tracing and Data Processing Using Trace Writers}
\label{app:tudo}
In this appendix, we give more insight on details that exceed the scope of this appendix's respective case study.

\subsubsection*{Additional Trace Writer Exercises}
\label{app:tudo:exercises}
Complementary to the respective case study's step-by-step instructions, in the toolchain's repository, there are also exercises that teach how to add a new trace writer to NVMain Each exercise comes with some base code and an example solution. To follow these exercises, go into the repository's submodule that is located at \texttt{repository\_root/simulator/nvmain} and checkout one of the following branches:
\begin{itemize}
    \item Trace-Writer-Tutorial-Ex1: Base for exercise 1 - Setting up a trace writer skeleton.
    \item Trace-Writer-Tutorial-Ex1-Solution: Example solution for exercise 1.
    \item Trace-Writer-Tutorial-Ex2: Base for exercise 2 - Put actual data to the trace.
    \item Trace-Writer-Tutorial-Ex2-Solution: Example solution for exercise 2.
\end{itemize}
Everything that we provide for the exercise can be found under\\
\texttt{repository\_root/simulator/nvmain/traceWriter/ESWeekTraceWriter/},\\
including the instructions for the exercise.

\subsection*{Heterogeneous Cache Simulations}\label{app:fau}
In this appendix, we provide a concrete example on how to run a gem5 simulation with hybrid caches enabled, as shown in \cref{sec:nvm-cache}.
As part of the repository, we provide a specific configuration named \texttt{fs\_hy.py}, where the instantiated caches are objects of the \texttt{HybridCache} class.
To perform a simulation with hybrid caches and a non-volatile main memory simulated in NVMain, the following command can be used:
\begin{lstlisting}[language=bash]
$ ./build/ARM/gem5.opt configs/example/fs_hy.py --bare-metal --kernel=path_to_unikraft_kernel --cpu-type=O3CPU --sys-clock=240MHz --cpu-clock=480MHz --machine-type=VExpress_GEM5_V2 --dtb-filename=system/arm/dt/armv8_gem5_v2_1cpu.dtb --mem-size=4GB --mem-type=NVMainMemory --nvmain-config=../nvmain/Config/PCM_ISSCC_2012_4GB.config --caches --l1d_size=32kB --l1i_size=32kB --l1d_assoc=4 --l1i_assoc=2 --l1d_nv_block_ratio=50
\end{lstlisting}
Whereas, the \texttt{l1d\_size} or \texttt{l1d\_assoc} are part of gem5's default \texttt{fs.py} configuration to set the size and associativity of the L1 data cache, we have added the \texttt{l1d\_nv\_block\_ratio} parameter.
This parameter can be set to any desired value between 0 and 100, setting the percentage of L1 data cache lines per set that are treated as being implemented in NVM technology, as explained in \cref{sec:nvm-cache}.
Vice versa, for the L1 instruction cache, these settings can be changed via the prefix \texttt{l1i} (instead of \texttt{l1d} in front of the parameter name.
\begin{lstlisting}[language=Python]
class L1Cache(HybridCache):
    ...
    data_read_latency = 2
    data_write_latency = 8
    vol_read_energy = 0.009
    non_vol_read_energy = 0.007
    vol_write_energy = 0.009
    non_vol_write_energy = 0.056
\end{lstlisting}
In addition to the latency settings, the class definitions in \texttt{configs/common/HybridCaches.py} also contains parameters for the dynamic energy consumption regarding read- and write accesses to the volatile and non-volatile cache section, respectively.
For experimental reasons, the parameters in this file can be set to any desired value.
\par
After performing a simulation, by default, the resulting statistics are dumped in \texttt{m5out/stats.txt}.
Here, among many other statistics, the number of CPU cycles, as well as the number of accesses to each cache section and the resulting dynamic energy consumption, according to the previously set energy parameters, are dumped as shown in the following:
\begin{lstlisting}
system.cpu.numCycles                ... # Number of cpu cycles simulated (Cycle)
...
system.cpu.dcache.noOfNonVolReads   ... # Number of reads to non-vol cache blocks (Count)
system.cpu.dcache.noOfVolReads      ... # Number of reads to vol cache blocks (Count)
system.cpu.dcache.noOfNonVolWrites  ... # Number of writes to non-vol cache blocks (Count)
system.cpu.dcache.noOfVolWrites     ... # Number of writes to vol cache blocks (Count)
system.cpu.dcache.dynEnergy         ... # Dynamic energy caused by cache accesses (nJ)
\end{lstlisting}
Using this statistic dump and the above mentioned command to run the simulation with different \texttt{l1d\_nv\_block\_ratio} settings, we can generate plots as shown in our case study.
\subsection*{Compute-in-Memory (CiM)}
This appendix provides code samples (\cref{kit/lst1}, \cref{kit/lst2}, and \cref{kit/lst3}) for implementing CiM application-level code and converting C++ code to CiM code with the help of the CiM API, as discussed in \cref{sec:nvm-cim}.
Additionally, in the following, it provides toolchain setup steps and utilizing CiM extention.

\begin{lstlisting}[language={[ANSI]C++}, label={kit/lst1}, caption={Simple vector-wise NAND operation implementation in a user application.},emph={cimModule},emphstyle=\color{cyan!50!blue}, emph={[2]CimModule},emphstyle={[2]\color{cyan}},
% numbers=right,numbersep=10pt,linewidth=0.9\linewidth
]
// Including CiM API
#include "cim_api.hpp"      
// ...
// Instantiating CiM object
CimModule cimModule;

// initializing CPU side arrays
uint8_t a[CimModule.ROW_SIZE] {0};
uint8_t b[CimModule.ROW_SIZE] {0};
uint8_t c[CimModule.ROW_SIZE] {0};

// initializing CiM rows
cimModule.copy_to_cim(0, (void *)&a[0]); // row0 <- a...
cimModule.copy_to_cim(1, (void *)&b[0]); // row1 <- b...
cimModule.copy_to_cim(2, (void *)&c[0]); // row2 <- c...

// Performing NAND operation
cimModule.AND({0,2});   // SA outputs = row0 & row2
cimModule.NOT_COND(1);   // row1 = ~SA outputs

// Copying results back to CPU side array
cimModule.copy_to_cim((void *)&b[0], 1); // b... <- row1
\end{lstlisting}

\begin{lstlisting}[language={[ANSI]C++}, label={kit/lst2}, caption={An example of using a ternary operator inside a for loop in C++.},emph={cimModule},emphstyle=\color{cyan!50!blue}, emph={[2]CimModule},emphstyle={[2]\color{cyan}},
% numbers=right,numbersep=10pt,linewidth=0.9\linewidth
]
for(size_t i=0; i< CimModule.ROW_SIZE; i++)
    a[i] = (b[i]==0x12) ? c[i] : d[i];
\end{lstlisting}
\begin{lstlisting}[language={[ANSI]C++}, label={kit/lst3}, caption={Converting the ternary operator to CiM code, along with unrolling the loop.},emph={cimModule},emphstyle=\color{cyan!50!blue}, emph={[2]CimModule},emphstyle={[2]\color{cyan}},
% numbers=right,numbersep=10pt,linewidth=0.9\linewidth
]
cimModule.copy_to_cim(0, (void *)&b[0]);            // row0 <- b...
cimModule.copy_to_cim(1, (void *)&CONSTx12[0]);     // row1 <- 0x12...
cimModule.copy_to_cim(2, (void *)&c[0]);            // row2 <- c...
cimModule.copy_to_cim(3, (void *)&d[0]);            // row3 <- d...
cimModule.copy_to_cim(4, (void *)&a[0]);            // row4 <- a...

cimModule.XOR({0,1})                    // (b[i]==0x12)
cimModule.NOT_COND(5, false, true);     // row5 <- 0xff if result is 0, otherwise 0x00
cimModule.NOT_COND(6, 5, true, false);  // row6 <- ~row5

cimModule.AND({2,5})                    // row7 <- row2 & row5
cimModule.COPY(7)

cimModule.AND({3,6})                    // row8 <- row3 & row6
cimModule.COPY(8)

cimModule.OR({7,8})                     // a <- row7 | row8
cimModule.COPY(4)
\end{lstlisting}
\subsubsection*{Minimal CiM Implementation from Scratch}
To understand how to implement and run full-system-based CiM simulations with gem5, and to build the required materials such as device drivers, please clone the toolchain's repository on GitHub~\cite{repo} using the `\texttt{-{}-branch cdnc}` flag. Then, open it in Visual Studio Code's development container mode. This will automatically download and initialize the necessary repositories.
The \texttt{cdnc} folder within the repository contains video tutorials on creating a custom disk image and transferring the necessary files into it using \texttt{QEMU}. Additionally, the \texttt{README} file includes a detailed step-by-step guide for running the provided examples.

To test and execute the provided examples and disk image, open Visual Studio Code, navigate to the `Terminal' tab, and select `Run Task'. Choose the task \texttt{Run skip-Steps1and2.sh} to download the required materials. Once the process is complete, select the \texttt{Run step3-testingWithGem5.sh} task to compile and execute gem5. If the setup is successful, you can access the gem5 guest terminal and run the examples as described in the mentioned \texttt{README} file.

\subsubsection*{Using the Provided CiM Extension}
To use the CiM extension, refer to the instructions in the Toolchain Setup appendix. After compiling gem5 with the appropriate flags (e.g., \cref{lst:flags}), navigate to the \texttt{simulator/gem5/tests/test-progs/CDNCcim} folder. This folder contains the application codes provided, as well as scripts to build and run the examples.

\end{document}